\documentclass[aps,prd,preprint,superscriptaddress,nofootinbib,floatfix]{revtex4-2}

\usepackage{amsmath,amssymb,bm}
\usepackage{graphicx,booktabs,multirow,placeins,xcolor}
\usepackage{hyperref}
\usepackage{booktabs}

\newcommand{\Phibar}{\Phi_{\rm bar}}
\newcommand{\Phimin}{\Phi_{\rm min}}
\newcommand{\Nbar}{N_{\rm bar}}
\newcommand{\Nturn}{N_{\rm turn}}
\newcommand{\gc}{\gamma_{\rm c}}
\newcommand{\Tstar}{T_\star}
\newcommand{\VK}{V_{\rm K}}
\newcommand{\Veff}{V_{\rm eff}}
\newcommand{\fcross}{f_{\rm cross}}
\newcommand{\ave}[1]{\langle #1\rangle}
\newcommand{\dd}{\mathrm{d}}
\newcommand{\ee}{\mathrm{e}}

\newcommand{\eq}[1]{Eq.~(\ref{#1})}
\newcommand{\eqs}[2]{Eqs.~(\ref{#1}) and (\ref{#2})}
\newcommand{\fref}[1]{FIG.~\ref{#1}}
\newcommand{\tref}[1]{TABLE~\ref{#1}}
\newcommand{\aref}[1]{Appendix~\ref{#1}}
\newcommand{\Fref}[1]{FIG.~\ref{#1}}
\newcommand{\Tref}[1]{TABLE~\ref{#1}}

\newcommand{\sref}[1]{Section \ref{#1}}
\usepackage{multirow}

\begin{document}

\title{Finite-Temperature Stabilization of a KKLT Modulus in Three-Dimensional Lattice Simulations}

\author{Saya Mahmoudi}
\email{sayamahmoudi@iyte.edu.tr}
\affiliation{Department of Physics, Izmir Institute of Technology,
Gülbahçe, Urla 35430, Izmir, Turkiye}

\author{Furkan Utku Biber}
\email{furkanutkubiber@gmail.com }
\affiliation{Department of Physics, Izmir Institute of Technology,
Gülbahçe, Urla 35430, Izmir, Turkiye}

\author{Hyukjung Kim}
\email{curi951007@gmail.com}
\affiliation{Lunit Inc., Gangnam-daero, Gangnam-gu, Seoul 06241, Republic of Korea}

\author{Heeseung Zoe}
\email{heeseungzoe@iyte.edu.tr}
\affiliation{Department of Physics, Izmir Institute of Technology,
Gülbahçe, Urla 35430, Izmir, Turkiye}

\date{\today}

\begin{abstract}
We investigate the cosmological stabilization of a KKLT modulus through three-dimensional Langevin lattice simulations incorporating an evolving thermal bath, dissipation, spatial gradients, and stochastic fluctuations. While the spatially averaged modulus closely follows homogeneous evolution away from the capture threshold, near-critical trajectories linger near the unstable potential barrier for approximately one e-fold, allowing long-wavelength fluctuations to undergo tachyonic amplification and drive localized regions across the barrier even as the mean field turns back toward the stabilized vacuum. Linear Langevin theory reproduces the fluctuation growth, while convergence tests identify a narrow stochastic layer in which the barrier-crossing fraction shows no systematic dependence on the discretization at a fixed effective distance from the threshold. These results demonstrate that capture of the spatially averaged modulus does not guarantee that every spatial region remains within the stabilizing barrier.
\end{abstract}

\maketitle

\newpage

\section{Introduction}
\label{sec:introduction}

String compactifications contain scalar moduli that determine the size and shape of the extra dimensions and the couplings of the four-dimensional effective theory~\cite{Douglas:2006es,Grana:2005jc}. In the KKLT construction~\cite{Kachru:2003aw}, the volume modulus is stabilized in a metastable vacuum separated from the decompactification direction by a potential barrier. However, the existence of this minimum does not guarantee cosmological stabilization. A modulus displaced from the minimum can acquire sufficient kinetic energy to cross the barrier and evolve toward decompactification, giving rise to the well-known overshoot problem~\cite{Brustein:1992nk,Barreiro:1998aj,Brustein:2004jp,Conlon:2022pnx}. Successful stabilization therefore depends on the initial conditions, the cosmological background, and the mechanisms that dissipate the modulus kinetic energy during its evolution~\cite{Huey:2000jx,Brustein:2004jp,Barreiro:2005ua,Kofman:2004yc,Apers:2022cyl,Apers:2024ffe,Brunelli:2025eif,Brunelli:2026ndh}.

Cosmological expansion and radiation affect modulus stabilization in several ways. A radiation background can prevent overshooting by sustaining Hubble damping as the modulus kinetic energy redshifts more rapidly than radiation~\cite{Brustein:2004jp}. At finite temperature, the field-dependent free energy of the thermal bath also modifies the effective potential and can remove the stabilizing minimum and barrier at sufficiently high temperatures~\cite{Buchmuller:2004xr,Buchmuller:2004tz,Anguelova:2007ex}. The consequences of thermal corrections depend on the compactification and stabilization mechanism, as illustrated by studies of KKLT-related constructions and large-volume scenarios~\cite{Anguelova:2009ht,Papineau:2008xf,Basiouris:2026eol}. Importantly, the disappearance of an instantaneous minimum does not necessarily imply decompactification. Time-dependent evolution can relax constraints inferred from the static thermal potential, although thermal corrections can substantially restrict the initial conditions allowing stabilization~\cite{Barreiro:2007hb}. The finite duration of reheating can further enlarge the range of successful initial conditions~\cite{Alam:2022rtt}, while the dynamics of gaugino condensates introduces additional restrictions~\cite{Papineau:2009tv}. These results emphasize that modulus stabilization is determined by cosmological evolution rather than by the instantaneous effective potential alone.

Spatial inhomogeneities introduce another aspect of the stabilization problem that is absent in homogeneous analyses. Previous lattice simulations of string moduli investigated oscillon formation and local overshooting, finding KKLT and LVS vacua stable against local overshooting within the parameter ranges studied~\cite{Antusch:2017flz}. A subsequent KKLT lattice study showed that oscillon formation requires initial conditions finely tuned near the potential maximum~\cite{Kasuya:2020szy}. These studies demonstrate the relevance of nonlinear spatial dynamics and the sensitivity of fluctuations to the initial field configuration. They also motivate examining the boundary between capture and overshoot, where the homogeneous trajectory approaches an unstable region of the potential and becomes particularly sensitive to perturbations.

A thermal environment adds dissipation and stochastic fluctuations to this spatial problem. Interactions with thermalized particles can generate friction in the scalar-field evolution~\cite{Yokoyama:2004pf,Yokoyama:2006wt,Nakayama:2008ks,Mukaida:2012qn,Bodeker:2006ij,Drewes:2014pfa,Banerjee:2025qyg}, accompanied by stochastic noise through the fluctuation--dissipation relation~\cite{Berera:1998gx,Bastero-Gil:2010dgy,Bastero-Gil:2012akf,Moss:2006gt,Bartrum:2014fla}. Near the capture threshold, the homogeneous modulus approaches the unstable barrier top and can remain there for an extended interval before turning back toward the stabilized vacuum. The negative curvature of the effective potential then amplifies long-wavelength fluctuations, allowing localized regions to cross the barrier even when the homogeneous trajectory is captured. The central question is therefore whether capture of the spatially averaged modulus is sufficient to ensure stabilization throughout space.

In this work, we investigate this question using three-dimensional Langevin lattice simulations of a KKLT modulus coupled to an evolving thermal bath. We adapt the Langevin lattice approach previously employed for thermal-inflation phase transitions~\cite{Kim:2026mal} to the finite-temperature modulus potential and implement it in \texttt{CosmoLattice}~\cite{Figueroa:2020rrl,Figueroa:2021yhd}. The simulations include Hubble expansion, spatial gradients, stochastic fluctuations, and energy exchange between the modulus and radiation. We parameterize the additional dissipation phenomenologically as $\eta=\gamma H$ and determine the homogeneous capture threshold independently. Linear Langevin theory provides a prediction for the fluctuation growth and local barrier-crossing fraction, while resolution and convergence tests assess the robustness of the lattice results.

We find that the spatially averaged modulus closely follows the homogeneous evolution away from the capture threshold. Near the threshold, however, the field remains close to the unstable barrier top for approximately one e-fold, during which long-wavelength fluctuations undergo substantial tachyonic amplification. As a result, localized regions cross the potential barrier even when the spatially averaged field initially turns back toward the stabilized vacuum. The linear Langevin calculation reproduces the fluctuation growth, and convergence tests identify a narrow stochastic layer in which the crossing fraction is stable when compared at a fixed effective distance from the threshold. Extended simulations show that some barrier-crossed regions continue toward larger field values, although their ultimate fate remains unresolved. These results demonstrate that homogeneous capture alone does not establish stabilization throughout space.

The paper is organized as follows. In \sref{sec:model}, we introduce the finite-temperature KKLT model and its coupling to the radiation bath. In \sref{sec:method}, we describe the stochastic lattice simulations and observables. In \sref{sec:results}, we present the homogeneous capture threshold, fluctuation amplification near the barrier, numerical convergence tests, and the evolution of barrier-crossed regions. In \sref{sec:discussion}, we discuss the physical implications and limitations of our results. The appendices provide the linear Langevin analysis, thermodynamic relations, numerical validation, and near-threshold discretization effects.

\section{Finite-temperature KKLT modulus}
\label{sec:model}

\subsection{Zero-temperature potential}
\label{sec:potential}

We consider the KKLT volume modulus $T=\sigma+\mathrm{i}\tau$, described by the K\"ahler potential $K=-3\ln(T+\bar T)$ and the superpotential $W=W_0+A\,\ee^{-aT}$~\cite{Kachru:2003aw}. Assuming that the axion $\tau$ is fixed at its minimum, the canonically normalized field is given by
\begin{equation}
\Phi=\sqrt{\tfrac32}\,\ln\sigma,\qquad \sigma=\ee^{\sqrt{2/3}\,\Phi}.
\label{eq:sigmaPhi}
\end{equation}
The resulting F-term potential, supplemented by an uplift contribution $d/\sigma^3$ motivated by anti-D3-branes, takes the form~\cite{Kachru:2003aw}
\begin{equation}
\VK(\Phi)=\frac{aA\,\ee^{-a\sigma}}{2\sigma^{2}}\Big[c\,aA\sigma\,\ee^{-a\sigma}+W_0+A\,\ee^{-a\sigma}\Big]+\frac{d}{\sigma^{3}},\qquad c=\frac13.
\label{eq:VK}
\end{equation}
The coefficient $c=1/3$ follows from the K\"ahler potential specified above.\footnote{For $K=-3\ln(T+\bar T)$ and $W=W_0+A\ee^{-aT}$, the F-term potential $\ee^K[K^{T\bar T}|D_TW|^2-3|W|^2]$ gives $c=\tfrac13$. In~\cite{Brustein:2004jp}, Eq.~(15) is printed with $\tfrac12$ instead; the numerical example given there (the positions of the minimum and barrier, the barrier height, and the quoted field intervals) is reproduced only with $c=\tfrac13$ once $W_0$ is tuned to the reported vacuum energy.} This potential admits a metastable vacuum separated from the decompactification direction by a barrier, whose role in cosmological modulus stabilization has been studied in~\cite{Brustein:2004jp}.

\subsection{Thermal bath and equations of motion}
\label{sec:bath}

We describe the radiation bath by the free-energy density~\cite{Barreiro:2007hb}
\begin{equation}
F(\Phi,T)=\big[a_0+a_2\,g^2(\Phi)\big]T^4,
\qquad
a_0=-\frac{\pi^2g_*}{90},\quad
a_2=r\,a_0,\quad
g^2(\Phi)=\frac{4\pi}{\sigma}
=4\pi\,\ee^{-\sqrt{2/3}\,\Phi},
\label{eq:F}
\end{equation}
where $T$ is the radiation temperature, $g_*$ counts the relativistic degrees of freedom, and $r$ parametrizes the interaction correction. The modulus dependence of the gauge coupling follows from the gauge kinetic function~\cite{Buchmuller:2004xr,Buchmuller:2004tz}. For $r<0$, the thermal contribution drives the modulus toward larger volume. The leading perturbative correction for an SU($N_c$) gauge plasma gives $r=-5N_c/(16\pi^2)\simeq-0.095$ for $N_c=3$~\cite{Kapusta:1979fh,Arnold:1994ps,Arnold:1994eb,Kajantie:2002wa}. Since perturbation theory is not reliable for gauge couplings of order unity, we treat $r$ as a phenomenological parameter. We choose $r=-0.2$ and examine the range $-0.3\le r\le-0.1$. We restrict the evolution to the region $1+rg^2(\Phi)>0$, where the radiation energy density is positive.

The radiation energy density is $\rho_r=3|a_0|(1+r\ave{g^2})T^4$, where $\ave{\cdot}$ denotes the spatial average. The finite-temperature effective potential is
\begin{equation}
\Veff(\Phi;T)=\VK(\Phi)+a_2\,g^2(\Phi)\,T^4.
\label{eq:Veff}
\end{equation}
The thermal force is the derivative of the free energy at fixed temperature, which equals the force at fixed entropy (\aref{app:thermo}). Since the thermal contribution is already included in $\rho_r$, it is not counted separately in the Friedmann equation. The thermodynamic consistency of this prescription is discussed in the same appendix.

We include thermal dissipation through $\eta=\gamma H$, where $H=\dot a/a$ is the Hubble rate and $\gamma$ is a dimensionless control parameter. The modulus obeys the Langevin equation
\begin{equation}
\begin{split}
&\ddot\Phi+(3H+\eta)\dot\Phi
-\frac{\nabla^2\Phi}{a^2}
+\VK'(\Phi)
-\sqrt{\tfrac23}\,a_2g^2(\Phi)T^4=\xi,\\
&\ave{\xi(\mathbf x,t)\xi(\mathbf x',t')}
=\frac{2\eta T}{a^3}
\delta^3(\mathbf x-\mathbf x')\delta(t-t'),
\end{split}
\label{eq:eom}
\end{equation}
where $a(t)$ is the scale factor, $\mathbf x$ denotes comoving coordinates, and $\xi$ is Gaussian white noise with zero mean, normalized by the fluctuation--dissipation relation~\cite{Bartrum:2014fla}. The cosmological expansion is determined by
\begin{equation}
3H^2=
\Big\langle
\frac12\dot\Phi^2
+\frac{(\nabla\Phi)^2}{2a^2}
+\VK(\Phi)
\Big\rangle+\rho_r.
\label{eq:friedmann}
\end{equation}

The bath temperature evolves with the expansion and the energy exchanged with the modulus. We use the comoving entropy $S_c=4|a_0|a^3(1+r\ave{g^2})T^3$, whose evolution is governed by
\begin{equation}
\frac{\dd S_c}{\dd t}
=\frac{a^3\eta}{T}
\left(
\ave{\dot\Phi^2}
-\frac{T}{a^3\Delta x^3}
\right),
\label{eq:entropy}
\end{equation}
where $\Delta x$ is the comoving lattice spacing. The two terms account for energy exchanged through dissipation and stochastic noise, respectively. For $\eta=0$, the comoving entropy is conserved.

\begin{table}[t]
\centering
\renewcommand{\arraystretch}{1.0}
\begin{tabular*}{0.8\linewidth}{@{\extracolsep{\fill}}lll@{}}
\toprule
\textbf{Category} & \textbf{Parameters} & \textbf{Values} \\
\midrule
\multirow{5}{*}{Model}
& $a,\,A,\,c$ & $0.1,\,1,\,1/3$ \\
& $W_0,\,d$ & $-0.0915782,\,0.00178913$ \\
& $\Phimin,\,\Phibar$ & $4.5179,\,4.9651$ \\
& $V_{\min},\,V_{\rm bar}$ & $-6.6\times10^{-14},\,5.450\times10^{-9}$ \\
& $m_{\min},\,|\VK''(\Phibar)|^{1/2}$ & $6.1\times10^{-4},\,2.4\times10^{-4}$ \\
\midrule
Thermal bath
& $g_*,\,r$ & $100,\,-0.2$ \\
\midrule
\multirow{4}{*}{Initial condition}
& $\Phi_i,\,\dot\Phi_i,\,\rho_{r,i}$ & $3.2,\,0,\,10^{-4}$ \\
& $T_i,\,H_i$ & $4.394\times10^{-2},\,6.32\times10^{-3}$ \\
& $\VK(\Phi_i)/V_{\rm bar}$ & $3.7\times10^3$ \\
& $\Tstar$ & $1.4239\times10^{-2}$ \\
\midrule
\multirow{2}{*}{Lattice simulation}
& $N_{\rm lat},\,L$ & $256,\,16000$ \\
& $\Delta x,\,\Delta t$ & $62.5,\,2$ \\
\bottomrule
\end{tabular*}
\caption{Model parameters, initial conditions, and lattice settings. The minimum and barrier quantities refer to the zero-temperature potential in \eq{eq:VK}, while $\Tstar$ denotes the temperature at which the barrier reappears.}
\label{tab:params}
\label{tab:full_parameters}
\end{table}

The model parameters and initial conditions are summarized in \tref{tab:params}. We choose $a=0.1$ and $A=1$ in the KKLT potential and adjust $W_0$ and $d$ to obtain a nearly Minkowski vacuum at $\sigma_{\min}=40$ ($\Phimin=4.518$). Here $\Phimin$ and $\Phibar$ denote the minimum and barrier positions of the zero-temperature potential, with corresponding energies $V_{\min}$ and $V_{\rm bar}$. The mass at the minimum is $m_{\min}^2=\VK''(\Phimin)$. We denote by $\Tstar$ the temperature at which the minimum and barrier reappear as the bath cools. The parameters are chosen to compress the KKLT hierarchy. In the KKLT construction, the supersymmetric minimum before the uplift lies where $W_0=-A\,\ee^{-a\sigma}\,(1+\tfrac23a\sigma)$, and its energy is $V_{\rm AdS}=-a^2A^2\ee^{-2a\sigma}/(6\sigma)$~\cite{Kachru:2003aw}. A controlled construction requires $\sigma\gg1$ and $a\sigma>1$, which is achieved by tuning the flux-induced constant to $|W_0|\ll1$~\cite{Kachru:2003aw}. Since the energies near the minimum, including the barrier height, are then suppressed by $\ee^{-2a\sigma}$, the stabilized vacuum and its barrier lie exponentially below the energies on the steep part of the potential, where the modulus can begin its evolution~\cite{Brustein:2004jp}. We refer to this exponential separation of scales, controlled by $a\sigma_{\min}$, as the KKLT hierarchy. For example, $W_0=-10^{-4}$, $A=1$, and $a=0.1$ give $\sigma\simeq113$, i.e.\ $a\sigma\simeq11$, in~\cite{Kachru:2003aw}, while the example of~\cite{Brustein:2004jp} has $a\sigma_{\min}\simeq32$ and an initial potential energy about $10^{29}$ times the barrier height. We compress this hierarchy by taking a much larger $|W_0|=0.092$, for which $a\sigma_{\min}=4$ after the uplift and $\VK(\Phi_i)/V_{\rm bar}=3.7\times10^3$. The initial Hubble rate then exceeds the modulus mass only by a factor of about ten, $H_i/m_{\min}\simeq10$, so that the whole evolution from the initial roll to the end of the hilltop dwell (\sref{sec:fluct}) spans about four e-folds and fits into a single three-dimensional lattice simulation. This compression makes the simulations computationally feasible; its consequences for our results are discussed in \sref{sec:discussion}.

\begin{figure}[!htbp]
\centering
\includegraphics[width=\linewidth]{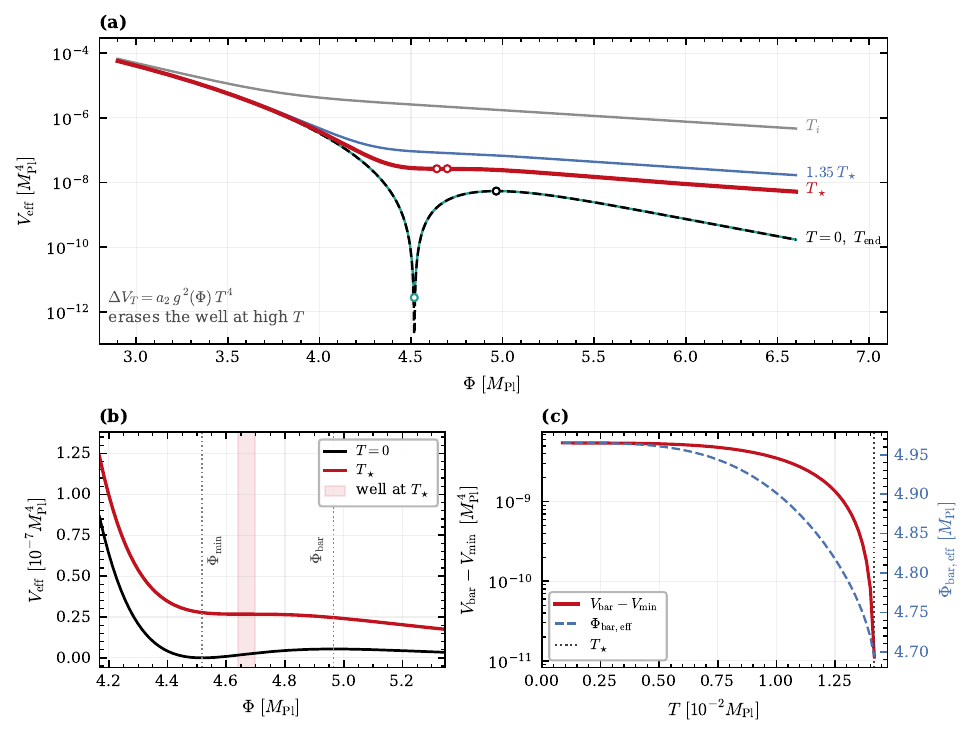}

\caption{Temperature dependence of the KKLT effective potential. (a) Potential at the indicated temperatures; circles mark extrema. (b) Stabilized region at $T=0$ and $\Tstar$; dotted lines mark the zero-temperature minimum and barrier. (c) Temperature-dependent barrier height and position.}
\label{fig:potential}
\end{figure}

\subsection{Barrier erasure, capture and overshoot}
\label{sec:capture}

\Fref{fig:potential} shows the temperature dependence of the effective potential for the parameters in \tref{tab:params}. At the initial temperature $T_i$, the thermal correction removes the local minimum and barrier of the zero-temperature potential, allowing the modulus to roll from $\Phi_i$ toward larger field values. As the radiation cools, the minimum and barrier reappear at the critical temperature $\Tstar$, at e-fold $N=\Nbar$, where $N\equiv\ln(a/a_i)$ measures expansion from the initial scale factor $a_i$. For the chosen parameters, $\Nbar\simeq1.08$ in the absence of dissipation, and increases slightly when dissipation is included because of energy transfer to the radiation bath. Below $\Tstar$, the minimum and barrier gradually approach their zero-temperature positions as the thermal contribution decreases.

The subsequent evolution depends on whether the modulus loses sufficient kinetic energy to remain within the potential well. If the field crosses the barrier, it continues toward larger volume rather than settling into the metastable vacuum. Dissipation can prevent this overshoot by slowing the field before it reaches the barrier. We define the \emph{capture threshold} $\gc$ as the critical value of the dissipation parameter $\gamma$ separating stabilization from overshoot in the deterministic homogeneous evolution. For $\gamma>\gc$, the modulus turns around before reaching the barrier and returns toward the stabilized vacuum, whereas for $\gamma<\gc$, it crosses the barrier and evolves toward decompactification. In the presence of spatial fluctuations, however, the evolution of the average field does not necessarily determine the outcome in every region. This distinction motivates the stochastic lattice analysis presented below.

\section{Lattice simulation}
\label{sec:method}

We evolve the modulus using \texttt{CosmoLattice}~\cite{Figueroa:2020rrl,Figueroa:2021yhd} on a periodic cubic lattice with $N_{\rm lat}^3$ sites, where $N_{\rm lat}$ is the number of sites along each side, $L$ is the comoving side length, and $\Delta x=L/N_{\rm lat}$ is the comoving lattice spacing. The scale factor evolves according to \eq{eq:friedmann}. The deterministic part of the field equation is integrated using the second-order velocity-Verlet scheme with the standard seven-point lattice Laplacian and periodic boundary conditions. Thermal friction and noise are incorporated through a symmetric splitting, with an Ornstein--Uhlenbeck update applied for half a time step before and after each deterministic step~\cite{Leimkuhler:2013jcp,Leimkuhler:2013amrx}. For fixed $a$, $T$, and $\eta$, the canonical momentum $\pi=a^3\dot\Phi$ at each lattice site follows an Ornstein--Uhlenbeck process~\cite{gardiner2009,lemons2003}. Over a time interval $\tau$, its exact update is
\begin{equation}
\pi\to\ee^{-\eta\tau}\pi+
\sqrt{\frac{a^3T(1-\ee^{-2\eta\tau})}{\Delta x^3}}\,G,
\label{eq:OU}
\end{equation}
where $G$ is an independent Gaussian random variable with zero mean and unit variance. The symmetric splitting preserves the fluctuation--dissipation relation for the stochastic update. The kinetic energy exchanged during each update, including the random work of the noise, is transferred to the radiation bath, as described by \eq{eq:entropy}.

The homogeneous initial conditions are $\Phi_i=3.2$ and $\dot\Phi_i=0$, with initial radiation energy density $\rho_{r,i}=10^{-4}$. Spatial fluctuations are initialized around this homogeneous background using the classical thermal spectra $\ave{|\delta\Phi_{\mathbf k}|^2}=T_i/(k^2+m_0^2)$ and $\ave{|\delta\pi_{\mathbf k}|^2}=T_i$, where $m_0^2=\max[\Veff''(\Phi_i;T_i),0]$, the initial scale factor is normalized to unity, and $\delta\pi_{\mathbf k}$ denotes the Fourier component of the momentum fluctuation. These spectra are also used to check the thermal initialization. To examine the dependence on the initial fluctuations, we perform comparison runs with vacuum-like spectra $\ave{|\delta\Phi_{\mathbf k}|^2}=1/(2\omega_k)$ and $\ave{|\delta\pi_{\mathbf k}|^2}=\omega_k/2$, where $\omega_k^2=k^2+m_0^2$. Separate runs with stochastic noise switched off are used to distinguish the effects of the initial fluctuations from those of noise during the evolution. The reference simulation uses $\Delta t/\Delta x=0.032$, while comparison runs include values up to $0.064$, within the stability range of the integration scheme. The comoving particle horizon remains smaller than half the box size throughout the expanding simulations.

Spatial fluctuations can be measured directly from the field values at individual lattice sites, without applying any spatial averaging or filtering. We refer to this as the \emph{unsmoothed field}, whose standard deviation is defined by $\sigma_{\rm raw}^2=\ave{(\Phi-\ave\Phi)^2}$. This quantity includes contributions from all Fourier modes resolved by the lattice and is therefore sensitive to the ultraviolet cutoff. For a classical thermal spectrum proportional to $T/k^2$, the variance behaves as $\sigma_{\rm raw}^2\simeq Tk_{\max}/(2\pi^2)$, where $k_{\max}$ denotes the lattice momentum cutoff. Consequently, the unsmoothed variance depends on the lattice spacing and is not a continuum observable~\cite{Aarts:1997kp,Bodeker:1995pp,Cassol-Seewald:2007oak}.

To characterize fluctuations on a specified spatial scale, we introduce the \emph{smoothed field} by applying a Gaussian filter in Fourier space, $\Phi_\ell(\mathbf k)=\ee^{-k^2\ell^2/2}\Phi(\mathbf k)$, where $\ell=\ell_{\rm phys}/a$ is the comoving smoothing length corresponding to the physical length $\ell_{\rm phys}$. The filter suppresses modes with wavelengths shorter than the chosen smoothing scale. We then define
\begin{equation}
\sigma_\ell^2=\ave{(\Phi_\ell-\ave\Phi)^2},\qquad
\fcross^{\,\ell}=\ave{\Theta\big(\Phi_\ell(\mathbf x)-\Phibar(T)\big)},
\label{eq:fcross}
\end{equation}
where $\sigma_\ell$ is the standard deviation of the smoothed field, $\Theta$ is the Heaviside step function, and $\fcross^{\,\ell}$ is the fraction of the volume in which the smoothed field exceeds the instantaneous barrier $\Phibar(T)$. Our primary choice is the Hubble length, $\ell_{\rm phys}=H^{-1}$, giving $\sigma_H$ and $\fcross^{\,H}$. We also consider the shorter thermal scale $\ell_{\rm phys}=3/T$, giving $\sigma_T$ and $\fcross^{\,T}$.

The Hubble filter defines a spatial observable, not a division between frozen and oscillating modes. Near the barrier, $|\Veff''|^{1/2}\simeq5.5H$, so some unstable modes with $H<k/a<|\Veff''|^{1/2}$ are suppressed by this filter. A barrier crossing of the smoothed field therefore indicates a local excursion, not necessarily irreversible escape. At $\Nbar$, the Hubble and thermal smoothing lengths span approximately $7$--$8$ and $1$ lattice cells, respectively, for $N_{\rm lat}=256$. The Hubble-smoothed field is resolved, whereas the thermal-scale observable is marginally resolved and is used as an upper envelope. The classical-field approximation applies for sufficiently occupied modes, roughly $k\lesssim T$. At fixed $L=16000$, $k_{\rm Nyq}/T_i=0.57$, $1.14$, and $1.71$ for $N_{\rm lat}=128$, $256$, and $384$.

\begin{figure}[!htbp]
\centering
\includegraphics[width=\linewidth]{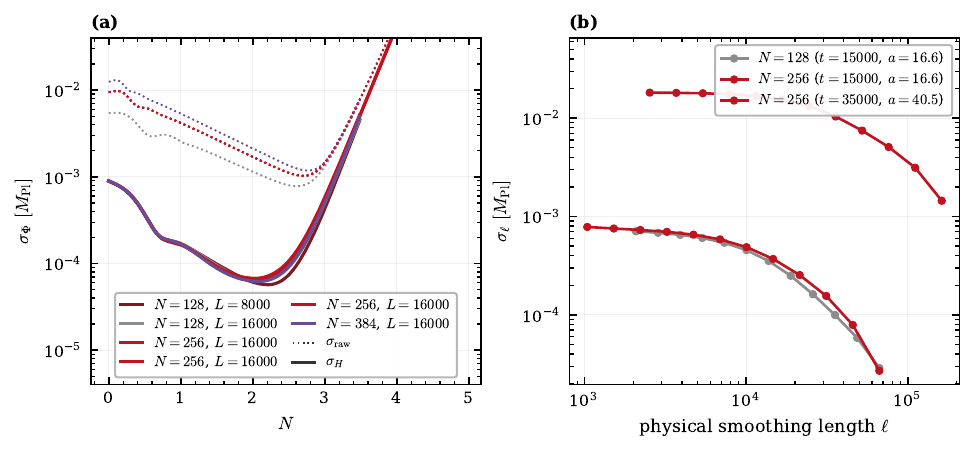}
\caption{Resolution and smoothing dependence of the field variance. (a) Unsmoothed variance $\sigma_{\rm raw}$ and Hubble-smoothed variance $\sigma_H$ for the runs at nominal $\gamma=0.50$. (b) Standard deviation $\sigma_\ell$ versus physical smoothing length.}
\label{fig:variance}
\end{figure}

Because the crossing fraction should describe the continuum field rather than the lattice, we check which of these fluctuation measures is independent of the lattice cutoff. \Fref{fig:variance}a compares $\sigma_{\rm raw}$ and $\sigma_H$ in runs at nominal $\gamma=0.50$ with $N_{\rm lat}=128$, $256$, and $384$ at $L=16000$, and with $N_{\rm lat}=128$ at $L=8000$, which has the same spacing as the reference lattice. Shortly after barrier reappearance ($N\simeq1.2$), $\sigma_{\rm raw}$ grows from $2.2\times10^{-3}$ to $3.5\times10^{-3}$ and $4.5\times10^{-3}$ as $N_{\rm lat}$ increases from $128$ to $256$ and $384$, and it takes the same value on the two lattices with $\Delta x=62.5$; it is therefore set by the lattice spacing, as expected for a variance dominated by modes near the cutoff. Over the same lattices, $\sigma_H$ changes by less than $5\%$, from $1.39\times10^{-4}$ to $1.32\times10^{-4}$. The larger differences in $\sigma_H$ at later times in \fref{fig:variance}a arise because these runs lie at different effective distances from the capture threshold (\sref{sec:convergence}), not from the cutoff. \Fref{fig:variance}b shows how $\sigma_\ell$ decreases as the physical smoothing length increases. We therefore use $\sigma_H$ and $\fcross^{\,H}$ as our primary observables.

We classify the outcome of each simulation using the spatially averaged field. A run is classified as \emph{overshot} if $\ave\Phi$ passes the instantaneous barrier after its reappearance. It is classified as \emph{captured} if $\ave{\dot\Phi}$ changes sign before the average field reaches the barrier and, at the end of the simulation, the average field cannot coast across the barrier. Runs in which the average field turns around but does not satisfy this condition at the end of the simulation are classified as \emph{undecided}. The detailed criterion and its limitations are given in \aref{app:validation}.

The simulation runs are organized into reference scan and spatial resolution tests (\tref{tab:runs_main}), convergence tests near $\gamma_c$ (\tref{tab:runs_critical}) and validation runs (\tref{tab:runs_validation}). Together, these sets comprise 122 simulations, including repeated realizations at the same $\gamma$. The numerical checks are summarized in \aref{app:validation}, and the near-threshold discretization effects in \aref{app:offsets}.

\begin{table*}[!t]
\centering
\renewcommand{\arraystretch}{1.0}
\setlength{\tabcolsep}{4pt}
\begin{tabular*}{\textwidth}{@{\extracolsep{\fill}}lccccrc@{}}
\toprule
\textbf{Set} & $N_{\rm lat}$ & \textbf{L} & $\bm{\Delta t}$ & $\bm{\gamma}$ & \textbf{Runs} & $\bm{t_{\text{max}}/10^4}$ \\
\midrule
Reference scan & $256$ & $16000$ & $2$ & $0 \text{--} 1$ $(11)$ & $17$ & $1.2 \text{--} 3$ \\
Resolution \& box size & $128, 384$ / $128$ & $16000$ / $8000$ & $2$ & $0.45, 0.50, 0.55$ & $9$ & $3$ \\
\bottomrule
\end{tabular*}
\caption{Reference parameter scan and spatial resolution tests. Numbers in parentheses indicate distinct values of $\gamma$. Unless stated otherwise, all runs start from $\Phi_i=3.2$, $\dot\Phi_i=0$, and $\rho_{r,i}=10^{-4}$ with the thermal initial spectrum, noise, and $\eta=\gamma H$. Two reference realizations at $\gamma=0.50$ are continued to $t=7\times10^4$.}
\label{tab:runs_main}
\end{table*}

\begin{table*}[!t]
\centering
\renewcommand{\arraystretch}{1.0}
\setlength{\tabcolsep}{4pt}
\begin{tabular*}{\textwidth}{@{\extracolsep{\fill}}lccccrc@{}}
\toprule
\textbf{Set} & $N_{\rm lat}$ & \textbf{L} & $\bm{\Delta t}$ & $\bm{\gamma}$ & \textbf{Runs} & $\bm{t_{\text{max}}/10^4}$ \\
\midrule
$\gamma$ scan & $128$ & $8000$ & $2$ & $0.49979 \text{--} 0.50049$ $(6)$& $18$ & $5.2$ \\
Time step & $128$ & $8000$ & $1, 0.5$ & $0.49943 \text{--} 0.49959$ $(3)$ & $9$ & $5.2$ \\
Spatial resolution & $192, 256$ & $8000$ & $2$ & $0.49824 \text{--} 0.49970$ $(3)$ & $6$ & $2.6 \text{--} 7$ \\
Homogeneous bisection & $16$ & $16000$ & $2, 1, 0.5, 0.25$ & near $\gc$ & $48$ & $6$ \\
\bottomrule
\end{tabular*}
\caption{Convergence tests near $\gamma_c$. Numbers in parentheses indicate distinct values of $\gamma$. The time-step set uses $\gamma=0.49959$ at $\Delta t=1$ and $\gamma=0.49943$, $0.49953$ at $\Delta t=0.5$. The spatial-resolution set uses $\gamma=0.49944$ at $N_{\rm lat}=192$ and $\gamma=0.49824$ at $N_{\rm lat}=256$, together with a pilot run at $\gamma=0.49970$ that fixes the offset for $\Delta x=31.25$; one realization at $\gamma=0.49824$ is continued to $t=7\times10^4$. The homogeneous bisection runs ($N_{\rm lat}=16$, without fluctuations or noise) determine the time-step offset of \aref{app:offsets}.}
\label{tab:runs_critical}
\end{table*}

\begin{table*}[!t]
\centering
\renewcommand{\arraystretch}{1.0}
\setlength{\tabcolsep}{4pt}
\begin{tabular*}{\textwidth}{@{\extracolsep{\fill}}lccccrc@{}}
\toprule
\textbf{Set} & $N_{\rm lat}$ & \textbf{L} & $\bm{\Delta t}$ & $\bm{\gamma}$ & \textbf{Runs} & $\bm{t_{\text{max}}/10^4}$ \\
\midrule
Decoupled bath ($\eta=0$) & \multirow{3}{*}{16 / 64} &  \multirow{3}{*}{$16000$ / $800$} & \multirow{3}{*}{$2, 1$} & \multirow{3}{*}{$0 \text{--} 0.55$} & \multirow{3}{*}{6} & \multirow{3}{*}{$0.5 \text{--} 3$} \\
Homogeneous background  & & &  &  & &  \\
Minkowski spacetime  &   & &  &  & &  \\ \hline
Noise off, vacuum initial spectrum, $\Delta t=1$ & \multirow{3}{*}{256} & \multirow{3}{*}{16000} & \multirow{3}{*}{2, 1} & \multirow{3}{*}{0.50} & \multirow{3}{*}{7} & \multirow{3}{*}{3} \\
$c=\tfrac12$, $r=-0.1$, $r=-0.3$ & & & & & & \\
$T^3$ friction law & & & & & & \\ \hline
Vacuum-initialized run & \multirow{2}{*}{256} & \multirow{2}{*}{16000} & \multirow{2}{*}{2} & $0.50$ & \multirow{2}{*}{2} & $7$ \\
Repeat with coarser output & & & & $0.52$ & & $3$ \\
\bottomrule
\end{tabular*}
\caption{Validation runs, model variants, and extensions. Validation sets verify thermodynamic consistency and zero-mode trajectories, while variants test the robustness of the mechanism. The decoupled-bath runs use $\Delta t=2$ and $1$; the homogeneous-background runs ($N_{\rm lat}=16$, without fluctuations or noise) use $\gamma=0.45$, $0.50$, and $0.55$; the Minkowski test ($N_{\rm lat}=64$, $L=800$) has no expansion, no potential, and constant $\eta$. Each variant at $\gamma=0.50$ changes one ingredient. The vacuum-initialized run is continued to $t=7\times10^4$, and the $\gamma=0.52$ run is a repeat with coarser output.}
\label{tab:runs_validation}
\end{table*}

\section{From zero-mode capture to stochastic escape}
\label{sec:results}

\subsection{Zero-mode dynamics and the capture threshold}
\label{sec:hom}
The evolution of the spatially homogeneous (zero-mode) background, $\phi(t)$, is governed by the deterministic equation of motion
\begin{equation}
\ddot\phi+(3H+\eta)\dot\phi+\VK'(\phi)-\sqrt{\tfrac23}\,a_2g^2(\phi)T^4=0,
\label{eq:mean_field}
\end{equation}
where the spatial gradients and stochastic noise are explicitly turned off. The scale factor and the temperature follow from the zero-mode forms of \eqs{eq:friedmann}{eq:entropy}, without the gradient and noise terms and with $\ave{g^2}=g^2(\phi)$. Here, $\eta = \gamma H$ parametrizes the additional friction. In this deterministic background, we define capture as the scenario where the field dissipates sufficient kinetic energy to turn around ($\dot\phi = 0$) before reaching the potential barrier maximum, ultimately settling into the local minimum (\aref{app:thermo}). Conversely, if the field retains enough energy to cross the barrier, it results in an overshoot. We first determine the capture threshold from \eq{eq:mean_field}, independently integrated using a fourth-order Runge--Kutta (RK4) method. For the model parameters in \tref{tab:params}, we obtain
\begin{equation}
\gc=0.499529,
\label{eq:gc}
\end{equation}
where $\gc$ is the critical dissipation parameter separating capture from overshoot in the homogeneous evolution. In particular, Hubble friction alone ($\gamma=0$) is insufficient to capture the modulus. This deterministic zero-mode threshold serves as the baseline against which we compare the three-dimensional lattice simulations with stochastic thermal fluctuations.

\begin{figure*}[!t]
\centering
\includegraphics[width=\textwidth]{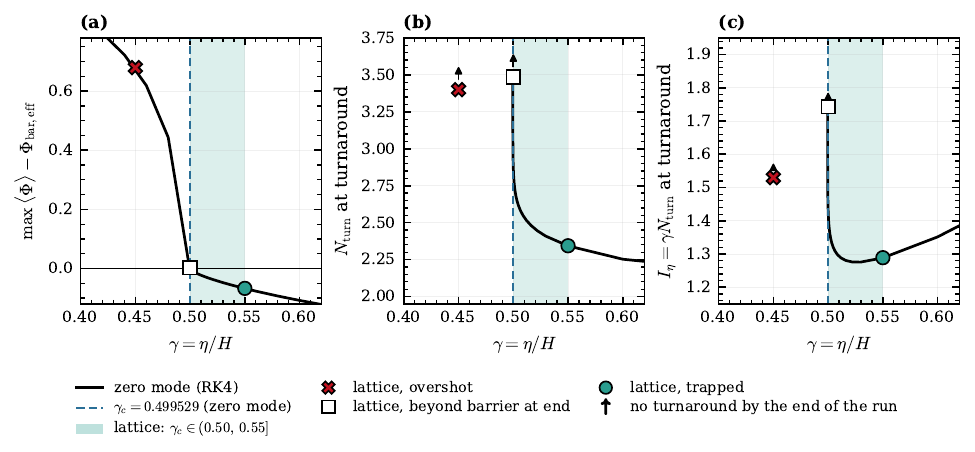}
\caption{Homogeneous modulus evolution and capture threshold. Solid curves show the independent zero-mode integration (RK4), and markers show homogeneous lattice runs ($N_{\rm lat}=16$) without fluctuations or noise. (a) Maximum of $\ave\Phi-\Phibar(T)$ up to $t=3\times10^4$. (b) Turnaround e-fold $\Nturn$. (c) Accumulated dissipation $I_\eta=\gamma\Nturn$. The dashed line marks $\gc=0.499529$, and the shaded band indicates the threshold bracket of the homogeneous lattice runs at $\Delta t=2$. Arrows denote runs without a turnaround by the end of the simulation.}
\label{fig:hom}
\end{figure*}

As $\gamma$ approaches $\gc$ from above, the modulus moves increasingly close to the potential barrier before turning around. We denote by $\Nturn$ the number of e-folds at which the field reverses its motion. Near the threshold, the turnaround is delayed logarithmically, and a numerical fit to our zero-mode solutions gives
\begin{equation}
\Nturn\simeq3.02-0.10\,\ln\!\Big(\frac{\gamma-\gc}{10^{-4}}\Big),
\qquad
10^{-5}\lesssim\gamma-\gc\lesssim3\times10^{-3}.
\label{eq:Nturnlog}
\end{equation}
The RK4 results are shown in \fref{fig:hom}, together with homogeneous lattice runs. The logarithmic dependence is characteristic of a trajectory approaching an unstable maximum.

The logarithmic delay can be understood by linearizing the zero-mode equation near the barrier maximum. The displacement contains a growing and a decaying mode, $\ee^{\lambda_\pm t}$, with rates $\lambda_\pm=-(3H+\eta)/2\pm\sqrt{(3H+\eta)^2/4+|\Veff''|}$. The coefficient of the growing mode vanishes at the separatrix and is proportional to $\gamma-\gc$ nearby, leading to a delay proportional to $\ln[1/(\gamma-\gc)]/(\lambda_+-\lambda_-)$. Near turnaround, $|\Veff''|/H^2\simeq27$ and $\eta\simeq0.5H$ give $H/(\lambda_+-\lambda_-)\simeq0.09$, consistent with the fitted coefficient $0.10$.

The accumulated dissipation up to the turnaround is characterized by $I_\eta=\int_0^{t_{\rm turn}}\eta\,dt=\gamma\Nturn$, where we have used $\eta=\gamma H$ and normalized the initial scale factor to unity. For $\gamma$ near $0.5$, we find $I_\eta\simeq1.4$. It remains above $1.27$ for all captured zero-mode trajectories considered here, indicating that dissipation acts over an appreciable interval before the field turns around.

We next compare the zero-mode results with the three-dimensional lattice simulations. In the reference $256^3$ scan, the barrier reappears within $0.006$ e-folds of the zero-mode prediction. For $\gamma\ge0.52$, the turnaround times agree within $2\times10^{-3}$ e-folds, to the time resolution of the stored simulation outputs. The gradient-energy fraction remains below $6.1\times10^{-4}$ during the capture process, and below $2.2\times10^{-3}$ in the $N_{\rm lat}=384$ runs. Thus, the spatially averaged field closely follows the homogeneous evolution even in the presence of thermal fluctuations.

\Tref{tab:zeromode} compares the barrier-reappearance time $\Nbar$, the turnaround time $\Nturn$, and the outcome of the zero-mode and lattice evolutions. The turnaround difference is defined as $\Delta\Nturn=\Nturn^{\rm lattice}-\Nturn^{\rm zero\ mode}$. The outcome ``turned'' indicates that the average field has reversed before the barrier but has not yet satisfied the capture criterion at the end of the simulation.

\begin{table*}[t]
\centering
\renewcommand{\arraystretch}{1.05}
\setlength{\tabcolsep}{4pt}
\begin{tabular*}{\textwidth}{@{\extracolsep{\fill}}cccccccl@{}}
\toprule
& \multicolumn{2}{c}{$\Nbar$}
& \multicolumn{3}{c}{$\Nturn$}
& \multicolumn{2}{c}{Outcome} \\
\cmidrule(lr){2-3}
\cmidrule(lr){4-6}
\cmidrule(lr){7-8}
$\gamma$ & Zero mode & Lattice & Zero mode & Lattice & $\Delta\Nturn$ & Zero mode & Lattice \\
\midrule
0.00 & 1.0840 & 1.088 & -- & -- & -- & overshot & overshot \\
0.25 & 1.1325 & 1.137 & -- & -- & -- & overshot & overshot \\
0.40 & 1.1562 & 1.158 & -- & -- & -- & overshot & overshot \\
0.45 & 1.1633 & 1.167 & -- & -- & -- & overshot & overshot \\
0.48 & 1.1675 & 1.168 & -- & -- & -- & overshot & overshot \\
0.50 & 1.1702 & 1.176 & 2.8648 & 3.028 & $+0.163$ & trapped & trapped \\
0.52 & 1.1728 & 1.177 & 2.4563 & 2.458 & $+0.002$ & trapped & turned \\
0.55 & 1.1767 & 1.178 & 2.3429 & 2.344 & $+0.001$ & trapped & trapped \\
0.60 & 1.1830 & 1.187 & 2.2527 & 2.253 & $+0.000$ & trapped & trapped \\
0.75 & 1.2003 & 1.200 & 2.1412 & 2.142 & $+0.001$ & trapped & trapped \\
1.00 & 1.2251 & 1.231 & 2.0935 & 2.095 & $+0.001$ & trapped & trapped \\
\bottomrule
\end{tabular*}

\caption{Comparison of the independent zero-mode integration and the reference $256^3$ lattice scan. $\Nbar$ and $\Nturn$ denote the e-folds at barrier reappearance and turnaround, respectively, while $\Delta\Nturn$ is their turnaround-time difference. Here $\gamma$ is the nominal value of the lattice runs. ``Turned'' indicates a trajectory that has reversed before the barrier but has not yet satisfied the capture criterion. At $\gamma=0.52$ the run ends on its second outward swing, and the coasting estimate of \aref{app:validation}, which neglects the restoring force of the potential, exceeds the distance to the barrier.}
\label{tab:zeromode}
\end{table*}

To test how the threshold depends on the model, we vary the thermal coefficient $r$, the potential coefficient $c$, and the dissipation law (\tref{tab:twins}). For the $c=\tfrac12$ variant, $W_0$ and $d$ are retuned to maintain $\sigma_{\min}=40$. For the dissipation, we consider a temperature-dependent friction law, $\eta=\gamma H_i(T/T_i)^3$, where $H_i$ and $T_i$ are the initial Hubble rate and temperature. Changing $r$ or $c$ shifts the zero-mode capture threshold by approximately $5\%$, while the $T^3$ friction law changes the normalization of $\gamma$ and produces a larger shift. In each variant, the lattice reproduces the zero-mode outcome at $\gamma=0.5$. The $T^3$-friction run exceeds the Friedmann residual tolerance (\aref{app:validation}) and is included for completeness.

\begin{table*}[t]
\centering
\renewcommand{\arraystretch}{1.05}
\setlength{\tabcolsep}{4pt}
\begin{tabular*}{\textwidth}{@{\extracolsep{\fill}}lcccc@{}}
\toprule
\textbf{Variant} & $\gc$ (zero mode) & \textbf{Outcome at $\gamma=0.5$} & $\Nbar$ & $\Nturn$ \\
& & Zero mode / lattice & & \\
\midrule
Reference ($c=\tfrac13$, $r=-0.2$) & 0.49953 & trapped / trapped & 1.170 / 1.176 & 2.865 / 3.028 \\
$r=-0.1$ & 0.47597 & trapped / trapped & 0.978 / 0.981 & 2.433 / 2.435 \\
$r=-0.3$ & 0.52504 & overshot / overshot & 1.279 / 1.282 & -- / -- \\
$c=\tfrac12$ & 0.52429 & overshot / overshot & 1.120 / 1.122 & -- / -- \\
$T^3$ friction law & 1.7165 & overshot / overshot & 1.122 / 1.129 & -- / -- \\
\bottomrule
\end{tabular*}
\caption{Parameter variation runs showing the dependence of the capture threshold on the model parameters and dissipation law. The values of $\Nbar$ and $\Nturn$ are given as zero mode / lattice.}
\label{tab:twins}
\end{table*}

The reference $256^3$ scan places the finite-time lattice capture threshold within the bracket
\begin{equation}
0.48 < \gc^{(3{\rm D})} \le 0.50,
\end{equation}
consistent with the deterministic zero-mode value $\gc = 0.499529$. While the reference realization at $\gamma=0.50$ is classified as captured at the end of its run ($N=3.49$), when the average field is still moving inward, two additional thermal realizations turn around, begin to move outward again at $N\simeq4.46$, and remain undecided at $N\simeq4.9$, and the vacuum-initialized run eventually crosses the barrier. This bracket should therefore be understood strictly as a finite-time classification of the spatially averaged field.

Across different lattice resolutions and time steps ($\Delta t$, $\Delta x$), the apparent threshold bracket shifts slightly around $\gamma = 0.50$. However, as shown in \aref{app:offsets}, these variations are consistent with numerical offsets of order $10^{-4}$--$10^{-3}$ arising from the finite time-step integration and the thermal energy density carried by resolved lattice modes, rather than with a shift of the background threshold caused by spatial gradients or nonlinear mode coupling. A finer parameter scan near the threshold ($N_{\rm lat}=128$, $L=8000$) narrows the transition to $0.49979 < \gc^{(3{\rm D})} \le 0.49994$, in quantitative agreement with the offset-predicted value of $0.49989$.

To compare runs across different discretizations on an equal footing, we assign each trajectory an effective dissipation parameter $\gamma_{\rm eff}$, defined as the zero-mode value of $\gamma$ that reproduces the lattice turnaround time $\Nturn$. Because no fluctuation observable enters this calibration (\aref{app:offsets}), $\gamma_{\rm eff} - \gc$ provides a consistent, discretization-independent measure of the distance from the capture threshold for the fluctuation analysis that follows.

Three related quantities are therefore used throughout. The nominal value $\gamma$ is the input dissipation parameter of a simulation and labels the runs in the tables and figures. The capture threshold $\gc=0.499529$ is fixed by the zero-mode evolution alone. For a lattice realization, the distance from the threshold is $\gamma_{\rm eff}-\gc$, which may differ from $\gamma-\gc$ by the numerical offsets of \aref{app:offsets}. For the zero-mode and linear Langevin calculations, which have no such offsets, $\gamma_{\rm eff}=\gamma$ and the distance is $\gamma-\gc$. Unless stated otherwise, values of $\gamma$ quoted for lattice runs are nominal.

\subsection{Thermal fluctuations and the hilltop dwell}
\label{sec:fluct}

A turnaround of the spatially averaged field $\ave\Phi$ does not guarantee that every local region remains inside the barrier. Close to the capture threshold, $\ave\Phi$ turns around near the barrier top, where $\Veff''<0$, rather than immediately rolling back into the potential well. It can remain in this region for approximately one e-fold, extending over $3.0\lesssim N\lesssim4.1$ in the extended runs at nominal $\gamma=0.50$ (\sref{sec:fate}). We refer to this interval as the \emph{hilltop dwell}. Its duration increases as the threshold is approached, consistent with the logarithmic turnaround delay in \eq{eq:Nturnlog}. Radiation dilution allows the potential energy to dominate during the dwell, while the proximity to the threshold determines how long the field remains near the top. The expansion therefore accelerates: in the extended thermal realizations, the total equation-of-state parameter $w=p_{\rm tot}/\rho_{\rm tot}$ stays below $-0.9$ for most of the dwell and reaches $w\simeq\ensuremath{-0.98}$. This phase is not slow roll, since $|\VK''|/\VK\simeq10$ at the barrier top.

The barrier crossing studied here is distinct from ordinary thermal activation across the full potential barrier. In the Kramers description, the escape rate is suppressed by $\exp(-\Delta E/T)$, with $\Delta E$ the barrier energy~\cite{Kramers:1940zz,Hanggi:1990zz}. For a homogeneous Hubble-sized region, a rough estimate gives $\Delta E/T\simeq(4\pi/3)H^{-3}V_{\rm bar}/T\simeq1.0\times10^8$ at $T=2.3\times10^{-3}$ and $H=4.6\times10^{-5}$. This suppresses thermal activation over the full barrier, but does not estimate the crossing of fluctuations around a homogeneous background already near the top. The dynamics also differs from the nearly flat-barrier regime of a Hawking--Moss transition~\cite{Hawking:1981fz}, since $|\Veff''|/H^2\simeq30$. Instead, the mean motion brings the field to the unstable region, where pre-existing fluctuations are amplified. The amplification is tachyonic: fluctuations grow because the effective mass squared is negative near the barrier top. Linearizing the field equation in this region gives the growing-mode rate
\begin{equation}
\lambda=-\frac{3H+\eta}{2}
+\sqrt{\frac{(3H+\eta)^2}{4}+|\Veff''|}
\simeq4.1H,
\end{equation}
where $\lambda$ is the positive growth rate for $H=\ensuremath{4.3\times10^{-5}}$ and $\eta=0.5H$~\cite{Felder:2000hj}.

We quantify the susceptibility to local barrier crossing through the dimensionless \emph{margin}
\begin{equation}
m_\ell=\frac{\Phibar(T)-\ave\Phi}{\sigma_\ell}\equiv\frac{d}{\sigma_\ell},
\label{eq:margin}
\end{equation}
which measures the distance $d=\Phibar(T)-\ave\Phi$ from the instantaneous barrier in units of the fluctuation amplitude $\sigma_\ell$ smoothed on the scale $\ell$ (\eq{eq:fcross}). As confirmed by our resolution tests (\fref{fig:variance}), the Hubble-smoothed amplitude $\sigma_H$ ($\ell = H^{-1}$) provides a cutoff-insensitive continuum observable, whereas the unsmoothed amplitude $\sigma_{\rm raw}$ varies with the lattice spacing. At barrier reappearance, for example, the linear Langevin calculation of \aref{app:linear} gives $\sigma_H = 1.4\times10^{-4}$ and $\sigma_{\rm raw} = 4.0\times10^{-3}$ for the reference lattice.

\begin{figure}[!htbp]
\centering
\includegraphics[width=\linewidth]{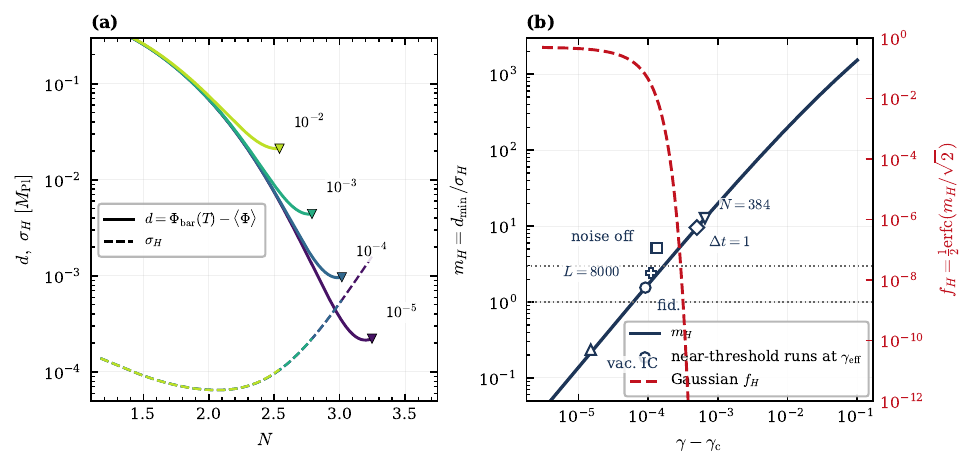}
\caption{Linear Langevin predictions near the capture threshold for the reference lattice. (a) Barrier distance $d$ (solid) and Hubble-smoothed standard deviation $\sigma_H$ (dashed) along zero-mode trajectories with $\gamma-\gc=10^{-5}$, $10^{-4}$, $10^{-3}$, and $10^{-2}$; triangles mark turnaround. (b) Margin $m_H$ at closest approach (left axis) and Gaussian crossing fraction $f_H$ (right axis) as functions of $\gamma-\gc$. Open symbols show the six lattice checks of \tref{tab:fluct}, all at nominal $\gamma=0.50$, each placed at its own $\gamma_{\rm eff}-\gc$; dotted lines mark $m_H=1$ and $3$.}
\label{fig:lintheory}
\end{figure}

On the zero-mode background, the linear Langevin calculation of \aref{app:linear} predicts $\sigma_H$ and the margin after $\gamma_{\rm eff}$ is fixed from turnaround, without fitting fluctuation data. For a Gaussian distribution, the Hubble-smoothed crossing fraction is
\begin{equation}
f_H=\frac12\,{\rm erfc}\!\left(\frac{m_H}{\sqrt2}\right),
\label{eq:fH}
\end{equation}
where $m_H=d/\sigma_H$ is the Hubble-scale margin and $\operatorname{erfc}$ is the complementary error function. The quantity $f_H$ estimates the lattice observable $\fcross^{\,H}$. It is negligible far above the threshold and increases as $\gamma\to\gc^+$ (\fref{fig:lintheory}). Before turnaround, the margin reaches its minimum $m_H\simeq3.6$, $1.7$, and $0.14$ for $\gamma-\gc=2\times10^{-4}$, $10^{-4}$, and $10^{-5}$, respectively.

We test these predictions against six near-threshold lattice runs that differ in noise, initial spectrum, time step, and resolution (open symbols in \fref{fig:lintheory}b). Because $\gamma_{\rm eff}$ is calibrated from $\Nturn$, agreement in $d$ is partly built in, and the fluctuation amplitude provides the independent test. Without any fit to fluctuation data, the linear theory reproduces $\sigma_H$ and $m_H$ to about $10\%$ over margins from below unity to above ten (\aref{app:linear} and \tref{tab:fluct}). The crossing fraction is tested over an ensemble of realizations in \sref{sec:convergence}.

\begin{figure}[!htbp]
\centering
\includegraphics[width=0.9\linewidth]{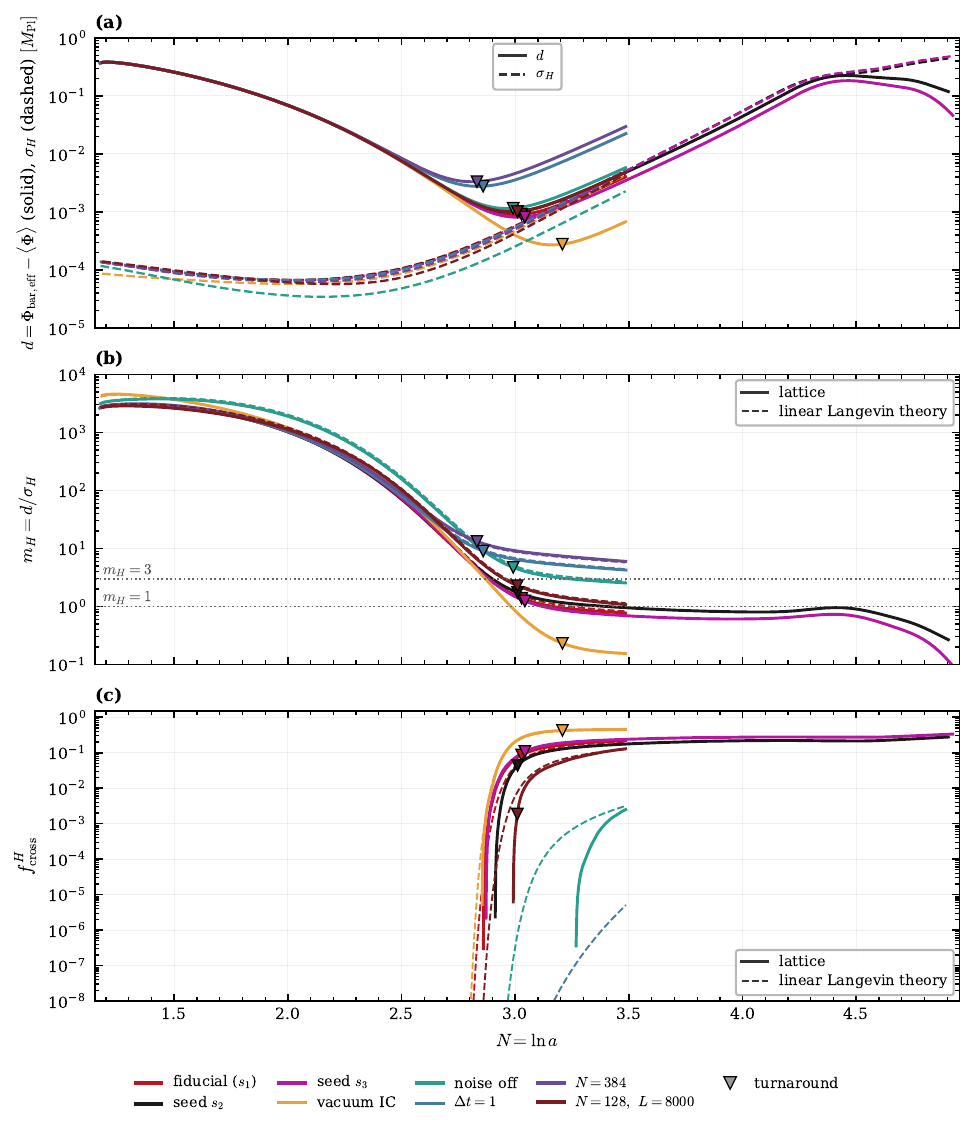}
\caption{Hilltop dwell in the near-threshold runs, including three extended realizations. (a) Barrier distance $d$ (solid) and Hubble-smoothed standard deviation $\sigma_H$ (dashed). (b) Margin $m_H$, with dotted lines at $1$ and $3$. (c) Hubble-smoothed crossing fraction. Triangles mark turnaround; thin dashed lines show the linear Langevin predictions continued beyond it. All runs have nominal $\gamma=0.50$; their effective distances $\gamma_{\rm eff}-\gc$ are $9.1\times10^{-5}$ ($s_1$), $\ensuremath{1.1\times10^{-4}}$ ($s_2$), $\ensuremath{7.9\times10^{-5}}$ ($s_3$), $1.3\times10^{-4}$ (noise off), $1.5\times10^{-5}$ (vacuum initial spectrum), $5.0\times10^{-4}$ ($\Delta t=1$), $6.5\times10^{-4}$ ($N_{\rm lat}=384$), and $1.1\times10^{-4}$ ($N_{\rm lat}=128$, $L=8000$), and each linear prediction is evaluated at the corresponding value. Fractions below one lattice site are omitted.}
\label{fig:dwell}
\end{figure}

The crossing fraction continues to increase even after the average field has turned around (\fref{fig:dwell}c). During the dwell, both the barrier distance $d$ and the fluctuation amplitude $\sigma_H$ grow, but their ratio $m_H=d/\sigma_H$ decreases only slowly. The Hubble-smoothed crossing fraction reaches $\ensuremath{2.2\times10^{-1}}$ in the reference run and $0.45$ in the vacuum-initialized run, which lies closer to its capture threshold; the effect therefore does not require a thermal initial spectrum. The same growth appears for every smoothing choice, although the fraction itself depends on the filter (\aref{app:validation}). By contrast, the runs at nominal $\gamma=0.52$ and $0.55$, which lie about $2\times10^{-2}$ and $5\times10^{-2}$ above the threshold, never have even one lattice site beyond the barrier. The continued linear theory reproduces $d$ and $\sigma_H$ through the weakly nonlinear stage of the dwell. Crossing fractions below approximately $10^{-2}$ are not statistically meaningful in a single realization, since the Hubble-smoothed field contains only about $3\times10^2$ independent cells at $N=3.5$ and $10^3$ at $N=4$.

We define the \emph{noise layer} in linear theory by a minimum Hubble margin $m_H<3$ during the hilltop dwell. Its estimated width above $\gc$ for the reference configuration is
\begin{equation}
\delta\gamma_{\rm noise}\simeq5\times10^{-4}.
\end{equation}
Evaluating the margin only through turnaround gives the smaller width $2\times10^{-4}$ (\tref{tab:lintable}). In linear theory, $d_{\min}\propto(\gamma-\gc)^{0.66}$ and $\sigma_H\propto(\gamma-\gc)^{-0.43}$, so $m_H\propto(\gamma-\gc)^{1.1}$. The estimated width scales roughly with the fluctuation amplitude at barrier reappearance and applies to $N_{\rm lat}=256$, $L=16000$, with thermal initial fluctuations. In the next subsection we determine the layer edge directly from the lattice crossing fraction.

\subsection{Width of the stochastic layer}
\label{sec:convergence}

\begin{figure*}[!t]
\centering
\includegraphics[width=\textwidth]{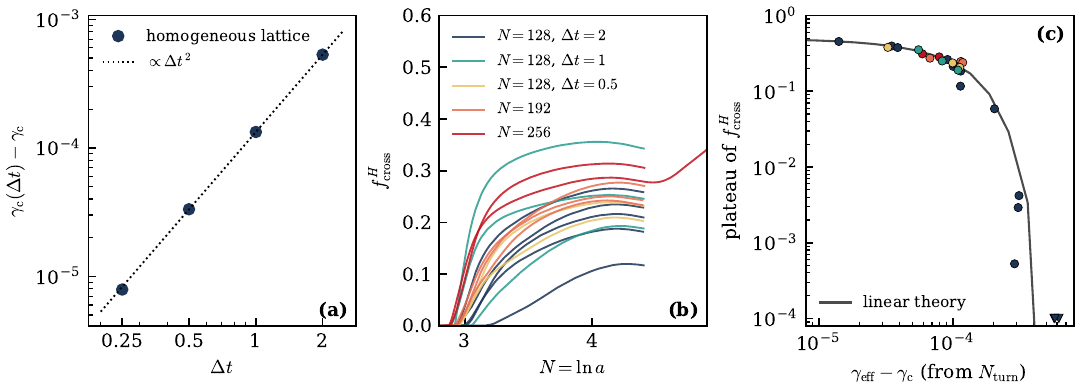}
\caption{Convergence tests near the capture threshold. (a) Homogeneous lattice threshold versus time step, with a dotted $\Delta t^2$ scaling. (b) Hubble-smoothed crossing fractions of the convergence runs with $5.5\times10^{-5}\le\gamma_{\rm eff}-\gc\le1.25\times10^{-4}$ for different time steps and lattice spacings. (c) Plateau crossing fraction versus $\gamma_{\rm eff}-\gc$, compared with linear Langevin theory. Triangles indicate fractions below $10^{-4}$.}
\label{fig:convergence}
\end{figure*}

At the same nominal $\gamma$, different discretizations can have different effective distances $\gamma_{\rm eff}-\gc$ from their capture thresholds. The six near-threshold checks span $6\times10^{-4}$ in $\gamma_{\rm eff}$, comparable to the noise-layer width. Accordingly, at nominal $\gamma=0.5$ the Hubble-smoothed crossing fraction ranges from $0.22$ in the reference run to below $10^{-10}$ at $\Delta t=1$ or $N_{\rm lat}=384$. We must compare equal effective distances, rather than equal nominal values, to test convergence.

We therefore perform a convergence ensemble in which the offset calibration of \aref{app:offsets} predicts $\gamma_{\rm eff}$ and the crossing fraction before each run (\tref{tab:convergence}). The box size $L=8000$ gives $\Delta x=62.5$ at $N_{\rm lat}=128$, retaining the reference spacing at lower cost. We vary $\Delta t=2,1,0.5$ at this spacing and use $\Delta x=41.7,31.25$ at $\Delta t=2$, targeting $\gamma_{\rm eff}-\gc\simeq9\times10^{-5}$. Six additional $\gamma$ values scan the layer at $\Delta t=2$, $\Delta x=62.5$; each configuration has two or three independent realizations.

The measured effective distances agree with the predictions. For the 23 runs that turned around, the difference between the measured and predicted $\gamma_{\rm eff}$ is $(-0.4\pm3.4)\times10^{-5}$, where the values denote the mean and standard deviation across the runs. This difference is comparable to the scatter between independent realizations. More importantly, at a fixed effective distance $\gamma_{\rm eff}-\gc$, the Hubble-smoothed crossing fraction shows no systematic dependence on the time step or lattice spacing (\fref{fig:convergence}b). We characterize the crossing fraction by its \emph{plateau}, defined as the median value during $3.9<N<4.4$, when it varies relatively slowly. For the 20 runs with a theoretical plateau above $0.05$, the ratio of the lattice plateau to the linear Langevin prediction is $0.96\pm0.13$, with no systematic trend in $\Delta t$ or $\Delta x$. Thus, the crossing fraction is primarily determined by the effective distance from the capture threshold, while much of its variation at fixed nominal $\gamma$ arises from numerical shifts of that threshold.

The ensemble also determines the layer edge (\fref{fig:convergence}c). The crossing plateau falls from $0.46$ at $\gamma_{\rm eff}-\gc=1.4\times10^{-5}$ to approximately $0.2$ at $10^{-4}$ and $0.06$ at $2\times10^{-4}$; at $3\times10^{-4}$ it lies between $5\times10^{-4}$ and $4\times10^{-3}$, and at $6\times10^{-4}$ it is below $10^{-15}$. Linear theory tracks fractions down to a few percent, but overestimates the small tail by factors of $2.6$--$32$ at $3\times10^{-4}$. In this tail the Gaussian approximation is least reliable and a single realization contains only a few independent barrier-crossing Hubble regions.

The runs end near $N=4.4$, so these plateaus are finite-time measurements. Taking a plateau above $10^{-3}$ as the lattice criterion gives an edge between $3\times10^{-4}$ and $6\times10^{-4}$ above the threshold, consistent with the linear noise-layer width derived using $m_H<3$. The two criteria differ, but identify the same narrow near-threshold regime.

\subsection{Evolution of barrier-crossed regions}
\label{sec:fate}

\begin{figure}[!htbp]
\centering
\includegraphics[width=\linewidth]{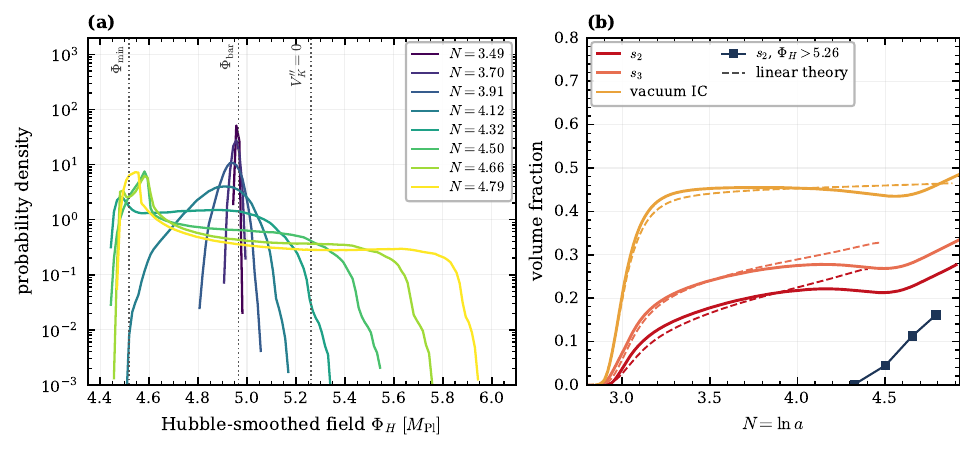}
\caption{Spatially inhomogeneous evolution after the hilltop dwell. (a) Distribution of the Hubble-smoothed field in $s_2$ (nominal $\gamma=0.50$, $\gamma_{\rm eff}-\gc=\ensuremath{1.1\times10^{-4}}$), with dotted lines marking the minimum, barrier, and outer inflection point at $\Phi=5.262$. (b) Barrier-crossing fractions for the three extended runs at nominal $\gamma=0.50$, whose effective distances are given in the text, and the fraction beyond the outer inflection point for $s_2$. Dashed curves show the continued linear Langevin predictions.}
\label{fig:fate}
\end{figure}

To determine whether barrier-crossed regions return to the stabilized vacuum or continue toward larger field values, we extend three near-threshold realizations at nominal $\gamma=0.50$ to $t=7\times10^4$. Two use thermal initial fluctuations and are labeled $s_2$ and $s_3$, the reference run being $s_1$; their effective distances from the capture threshold are $\gamma_{\rm eff}-\gc=\ensuremath{1.1\times10^{-4}}$ and $\ensuremath{7.9\times10^{-5}}$. The third starts from vacuum fluctuations, with $\gamma_{\rm eff}-\gc=\ensuremath{1.5\times10^{-5}}$, and is discussed separately because it belongs to a different initial ensemble.

During the hilltop dwell, the Hubble-smoothed field distribution broadens around the barrier. The crossing fraction reaches a plateau over $3.9\lesssim N\lesssim4.4$ and then increases by a further 5--30\% by $N\simeq4.9$. The median plateau fractions are \ensuremath{0.22}\ and \ensuremath{0.28}\ for the two thermal realizations and \ensuremath{0.45} for vacuum initial fluctuations. The continued linear theory follows these plateaus within $25\%$, but is systematically high as the field distribution becomes non-Gaussian.

\begin{figure*}[!t]
\centering
\includegraphics[width=\textwidth]{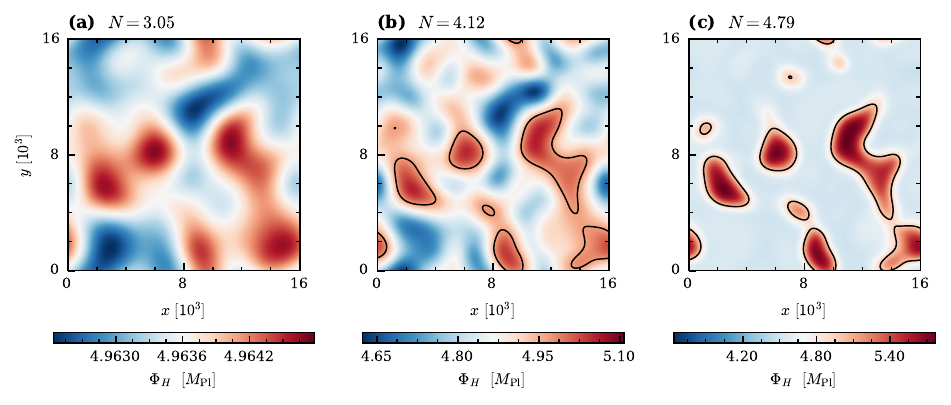}
\caption{Mid-plane slices of the Hubble-smoothed field $\Phi_H$ in the extended thermal realization $s_2$ (nominal $\gamma=0.50$, $\gamma_{\rm eff}-\gc=\ensuremath{1.1\times10^{-4}}$). Panels show (a) near turnaround, (b) the end of the hilltop dwell, and (c) $N=4.79$. Black contours mark $\Phi_H=\Phibar(T)$. In (a), the slice has no barrier-crossed region, although 7\% of the total volume lies beyond the barrier. Each panel uses its own color scale.}
\label{fig:slices}
\end{figure*}

After the homogeneous background leaves the tachyonic region, the distribution develops a peak near the KKLT minimum and a broad tail beyond the barrier (\fref{fig:fate}a). Most of the volume relaxes toward the minimum, while barrier-crossed regions continue toward larger $\Phi$. The spatial slices in \fref{fig:slices} show that these regions largely retain their shapes as the field inside them moves outward. Thus, a local escape channel opens even when the homogeneous background initially turns back.

To distinguish regions that have just crossed the barrier from those moving farther toward decompactification, we introduce a conservative runaway criterion based on the outer inflection point of the zero-temperature potential. This point is defined by $\VK''=0$ at $\Phi=5.262$, beyond which the outward potential force becomes weaker. Since $\VK'<0$ beyond $\Phibar$, the potential force favors increasing $\Phi$ throughout this region. The volume fractions beyond the outer inflection point are $\ensuremath{0.16}$ for $s_2$ and $\ensuremath{0.35}$ for the vacuum-initialized realization at $N=\ensuremath{4.8}$. Both fractions are still increasing at that time.

The vacuum-initialized realization shows that turnaround of the spatially averaged field need not imply its lasting capture. Although $\ave\Phi$ turns at $N=3.208$, it subsequently moves outward and reaches $\ensuremath{5.108}>\Phibar$ by $N\simeq5.0$, while the zero mode at the same $\gamma_{\rm eff}$ remains captured. In $s_2$, $\ave\Phi$ also begins to increase again from $N\simeq4.46$. Within the stochastic layer, local and averaged outcomes can therefore differ at late times.

The late-time evolution is less well controlled than the dwell. The interface width between crossed and uncrossed regions, estimated by $|\Veff''|^{-1/2}$, spans about one lattice cell at the reference spacing $\Delta x=62.5$ while the crossing fraction approaches its plateau, and only half a cell by $N=4.8$. A run at $\Delta x=31.25$ resolves the interface by two to three cells at the same times and shows the same escape behavior, with $23\%$ of the volume beyond the outer inflection point at $N=4.8$ (\aref{app:validation}). In addition, the gradient-energy fraction $\Omega_{\rm grad}$, the ratio of spatial gradient energy to the total energy density, reaches $0.08$--$0.10$ near the end. The Friedmann equation sourced by spatial averages, adequate while $\sigma_H\lesssim0.1$ ($N\lesssim4.1$), becomes less reliable as $\sigma_H$ grows to approximately $0.4$ at $N\simeq4.8$ (\sref{sec:discussion}). These simulations therefore establish the onset of spatially inhomogeneous escape, while the final fate of the escaping regions remains beyond the present setup.

\FloatBarrier

\section{Discussion}
\label{sec:discussion}

Our simulations reveal a distinction between the capture of the spatially averaged modulus and the evolution of individual spatial regions. Away from the capture threshold, the spatial average closely follows the homogeneous trajectory, whose critical dissipation parameter is $\gc=0.499529$ for the reference model. Near the threshold, however, the field lingers near the unstable barrier top for approximately one e-fold, allowing long-wavelength fluctuations to undergo tachyonic amplification and drive localized regions across the barrier. Linear Langevin theory reproduces the fluctuation growth, while the lattice crossing fraction shows no systematic dependence on the discretization when compared at a fixed effective distance from the threshold. The ratio of the lattice crossing plateau to the linear prediction is $0.96\pm0.13$. The stochastic layer extends $(3$--$6)\times10^{-4}$ above the threshold, consistent with the linear estimate of $5\times10^{-4}$. Extended simulations further show that some barrier-crossed regions continue toward larger field values, although their ultimate fate remains unresolved.

These quantitative results depend on the model parameters and should not be regarded as universal predictions of KKLT stabilization. Varying the thermal coefficient $r$ or the potential coefficient $c$ shifts the homogeneous capture threshold by approximately $5\%$, while a temperature-dependent friction law produces a larger shift (\sref{sec:hom}). Our simulations also employ a compressed KKLT potential with $a\sigma_{\min}=4$, compared with approximately $32$ in~\cite{Brustein:2004jp}. A larger hierarchy could substantially alter the modulus evolution before the stabilizing barrier reappears. Furthermore, the dissipation law $\eta=\gamma H$ is phenomenological, and microscopic damping of gravitationally coupled moduli may be much weaker than the Hubble rate~\cite{Bodeker:2006ij}. These assumptions affect the capture threshold, the width of the stochastic layer, and the fraction of barrier-crossed regions. Nevertheless, the underlying mechanism is more general: a field lingering near an unstable potential maximum can amplify long-wavelength fluctuations and develop spatially different outcomes.

The numerical treatment introduces additional limitations. The lattice has a finite ultraviolet cutoff characterized by the Nyquist wavenumber $k_{\rm Nyq}=\pi/\Delta x$ along each spatial direction, with $k_{\rm Nyq}/T_i=1.14$ for the reference lattice. Our classical stochastic model is not a cutoff-independent effective theory with counterterms. Although spatial smoothing reduces the cutoff sensitivity of the measured observables, it does not eliminate ultraviolet effects on the evolution itself. In particular, thermalized lattice modes shift the capture threshold, motivating comparisons at the same effective distance from the threshold. The correction to the mean force from unsmoothed fluctuations, $\tfrac12\Veff'''\sigma_{\rm raw}^2$, is estimated to be $1$--$4\%$ of $\Veff'$ near the barrier, corresponding to $\Delta\gamma\sim5\times10^{-6}$. The classical-field approximation also requires sufficiently occupied modes, roughly $k/a\lesssim T$. In the extended runs, $T/|\Veff''|^{1/2}$ ranges from $4$ to $6$, indicating substantial but not parametrically large occupation numbers for the unstable modes.

We treat the radiation as a homogeneous thermal bath and determine the cosmological expansion from spatially averaged energy densities. The homogeneous-bath approximation is not justified by rapid thermal diffusion across Hubble-sized regions. Rather, its use during the weakly inhomogeneous evolution is supported by the estimated smallness of temperature perturbations induced by modulus fluctuations. Near the barrier, the fixed-local-entropy estimate gives $\delta T/T\simeq1.2\times10^{-2}\,\delta\Phi$, corresponding to a correction to the thermal force of about $1\%$ of $\VK'$ and a change in the effective potential curvature of approximately $10^{-5}$ per unit $\delta\Phi$. These estimates support the approximations while $\sigma_H\lesssim0.1$, but they become less reliable as strong spatial inhomogeneities develop. At late times, the interfaces between barrier-crossed and uncrossed regions are only marginally resolved, and the four realizations followed into this regime are insufficient to determine ensemble probabilities. In addition, the axion $\tau$ is held at its minimum, although its fluctuations and couplings could modify the effective barrier. Our conclusions are therefore restricted to the single-field stochastic model and the regime in which its approximations remain controlled.

Our findings highlight an important distinction between homogeneous capture and spatially resolved modulus stabilization. The homogeneous capture criterion does not by itself guarantee that the modulus remains stabilized throughout space, since fluctuations amplified during the hilltop dwell can drive localized regions across the potential barrier. Whether these regions eventually return to the stabilized vacuum or continue toward decompactification remains an open question. In a companion paper currently in preparation, we investigate their spatial structure, connectivity, and interface dynamics using larger lattices and ensembles at different distances from the threshold. Establishing their ultimate fate will require a more complete treatment of inhomogeneous gravitational dynamics. Incorporating a dynamical axion, microscopic dissipation, and a less compressed KKLT hierarchy would further clarify the broader applicability of the mechanism.

\acknowledgments
The authors thank Gürcan Coşgel, Atakan Tuçel, Alpcan Bahar and Ahmet Şekerci for helpful discussions.
SM and HZ carried out part of this work during their stay at IBS-CTPU and would like to thank Kiwoon Choi, Sang Hui Im and Won Sang Cho for their warm hospitality. In this work we use CosmoLattice~\cite{Figueroa:2021yhd,Baeza-Ballesteros:2026uao,Florio:2026vde},
based on the lattice-cosmology techniques presented 
in~\cite{Figueroa:2020rrl,Baeza-Ballesteros:2025tme}.

\appendix

\section{Linear Langevin theory}
\label{app:linear}

We separate the modulus field into a homogeneous background and spatial fluctuations, $\Phi(\mathbf x,t)=\phi(t)+\delta\Phi(\mathbf x,t)$. The background quantities $\phi$, $a$, $H$, and $T$ obey the zero-mode form of \eqs{eq:eom}{eq:entropy}, obtained by setting $\nabla^2\Phi=0$, $\xi=0$, and $\ave{g^2}=g^2(\phi)$. We assume that the temperature does not respond to the fluctuations and neglect their back-reaction on the background. The fluctuation energy is small compared with the radiation energy, $\rho_{\rm fluct}/\rho_r\sim10^{-3}$, as discussed in \aref{app:offsets}. Expanding the field equation to first order in $\delta\Phi$, each comoving Fourier mode satisfies
\begin{equation}
\delta\ddot\Phi_{\mathbf k}+(3H+\eta)\,\delta\dot\Phi_{\mathbf k}+\omega_k^2\,\delta\Phi_{\mathbf k}=\xi_{\mathbf k},\qquad
\omega_k^2=\frac{k^2}{a^2}+M^2,\qquad
M^2=\VK''(\phi)+\tfrac23\,a_2\,g^2(\phi)\,T^4,
\label{eq:linmode}
\end{equation}
where $\omega_k^2$ is the squared frequency of the mode and $M^2=\Veff''(\phi,T)$ is the curvature of the effective potential at fixed temperature. The stochastic source $\xi_{\mathbf k}$ represents thermal noise and has the correlation
\[
\ave{\xi_{\mathbf k}(t)\xi_{\mathbf k'}^*(t')}=
\frac{2\eta T}{a^3}(2\pi)^3
\delta^3(\mathbf k-\mathbf k')\delta(t-t').
\]
The factor $2\eta T/a^3$ follows from the fluctuation-dissipation relation used in the lattice simulations.

For the linear Langevin system, the evolution can be expressed in terms of three equal-time correlators~\cite{gardiner2009,lemons2003}: the field variance $P_{FF}=\ave{|\delta\Phi_{\mathbf k}|^2}'$, the field--velocity correlation $P_{F\pi}=\mathrm{Re}\ave{\delta\Phi_{\mathbf k}\delta\dot\Phi_{\mathbf k}^*}'$, and the velocity variance $P_{\pi\pi}=\ave{|\delta\dot\Phi_{\mathbf k}|^2}'$. The prime removes the Fourier volume factor $(2\pi)^3\delta^3(0)$. Here the subscript $\pi$ denotes the field velocity $\delta\dot\Phi$, not the canonical lattice momentum $a^3\dot\Phi$. These correlators obey the closed system
\begin{align}
\dot P_{FF}&=2P_{F\pi},\nonumber\\
\dot P_{F\pi}&=P_{\pi\pi}-(3H+\eta)P_{F\pi}-\omega_k^2P_{FF},\label{eq:linP}\\
\dot P_{\pi\pi}&=-2(3H+\eta)P_{\pi\pi}-2\omega_k^2P_{F\pi}+\frac{2\eta T}{a^3}.\nonumber
\end{align}
The last term represents the continuous injection of fluctuations by the thermal noise. Since these equations are linear in the correlators, their evolution is determined entirely by the homogeneous background and the initial fluctuation spectrum.

We use the same initial spectra as in the lattice simulations. For thermal initial fluctuations, $P_{FF}=T_i/(k^2+m_0^2)$, $P_{F\pi}=0$, and $P_{\pi\pi}=T_i$, where $m_0$ is the initial mass parameter entering the thermal spectrum. For vacuum initial fluctuations, $P_{FF}=1/(2\omega_k)$, $P_{F\pi}=0$, and $P_{\pi\pi}=\omega_k/2$, with $\omega_k$ evaluated at the initial time. The mode correlators and homogeneous background are integrated together using RK4 with $\Delta t=1$.

To compare the linear calculation with the lattice observables, we compute the variance of the field smoothed over a comoving length $\ell$:
\begin{equation}
\sigma_\ell^2(t)=\int\frac{k^3\,\dd\ln k}{2\pi^2}\;\ee^{-k^2\ell^2}\,P_{FF}(k,t),\qquad
\ell_H=\frac1{aH},\quad\ell_T=\frac3{aT}.
\label{eq:sigmal}
\end{equation}
The exponential factor is the Gaussian smoothing filter. The lengths $\ell_H$ and $\ell_T$ correspond to the physical Hubble radius $H^{-1}$ and the thermal smoothing scale $3/T$, respectively. We denote their standard deviations by $\sigma_H$ and $\sigma_T$, while $\sigma_{\rm raw}$ is obtained by removing the filter. The integral is evaluated on 240 logarithmically spaced momentum shells between the infrared cutoff $k_{\rm IR}=2\pi/L$ and the largest lattice momentum $\sqrt3\,k_{\rm Nyq}$, where $k_{\rm Nyq}=\pi/\Delta x$. This spherical approximation to the cubic Brillouin zone also neglects the lattice dispersion relation. For an equilibrium spectrum, it changes $\sigma_H$, $\sigma_T$, and $\sigma_{\rm raw}$ by $-3.5\%$, $-2\%$, and $+4\%$, respectively. The effect is most important for the unsmoothed variance, which receives substantial contributions from modes near the lattice cutoff.

We characterize the probability of barrier crossing by the margin $m_\ell=d/\sigma_\ell$, where $d=\Phibar(T)-\phi$ is the distance of the homogeneous field from the instantaneous barrier. For the Hubble-smoothed field, the minimum margin before turnaround is $m_H^{\min}=\min_{\Nbar\le N\le\Nturn}[d(N)/\sigma_H(N)]$. The corresponding Gaussian crossing estimate is $f_H=\tfrac12{\rm erfc}(m_H^{\min}/\sqrt2)$, as defined in \eq{eq:fH}, while $f_T$ is obtained using the thermal smoothing scale. Here $d_{\min}$ denotes the minimum barrier distance over the same interval. The minima of $d$ and $d/\sigma_H$ need not occur at the same time.

\tref{tab:lintable} reports the linear predictions versus $\gamma-\gc$. As the zero mode lingers near the barrier top, $M^2<0$ amplifies long-wavelength fluctuations while the closest barrier distance decreases. The resulting margin falls from about $20$ at $\gamma-\gc=10^{-3}$ to below unity for $\gamma-\gc\lesssim3\times10^{-5}$. This makes crossing more likely but also limits the validity of a Gaussian approximation.

When fluctuations reach the nonlinear part of the potential, $f_H$ is only a linear-theory estimate. The shorter thermal filter includes more cutoff-scale modes, so $f_T$ is used only as an upper envelope. Direct lattice checks are given in \tref{tab:fluct}.

The six lattice checks in \tref{tab:fluct} are evaluated at $\gamma_{\rm eff}$ determined by turnaround. This table uses $\sigma_H^{\max}$, the maximum standard deviation before turnaround, and $d$ evaluated at turnaround, so its margin $m_H=d/\sigma_H^{\max}$ differs slightly from the minimum of $d/\sigma_H$ in \tref{tab:lintable}. Since no fluctuation observable determines $\gamma_{\rm eff}$, the comparison of fluctuation amplitudes is independent of the spatially averaged field matching. The six runs are the reference run, a run without noise, a run with vacuum initial fluctuations, $\Delta t=1$, $N_{\rm lat}=384$, and $N_{\rm lat}=128$ with $L=8000$. The linear theory underestimates $\sigma_H$ by 5--10\% in four runs, agrees within 1\% for vacuum initial conditions, and overestimates it by 4\% for $N_{\rm lat}=384$. About $3.5\%$ of the low bias arises from treating the Brillouin zone as spherical in \eq{eq:sigmal}. Since each check is a single realization, the seed-to-seed scatter of $\gamma_{\rm eff}$, $\ensuremath{3\times10^{-5}}$, can change $m_H$ by about $30\%$ near $\gamma_{\rm eff}-\gc=10^{-4}$.

\begin{table*}[t]
\centering
\renewcommand{\arraystretch}{1.0}
\setlength{\tabcolsep}{4pt}
\begin{tabular*}{\textwidth}{@{\extracolsep{\fill}}ccccccc@{}}
\toprule
$\gamma-\gc$ & $\Nturn$ & $d_{\min}$ & $\sigma_H$ & $m_H$ & $f_H$ & $f_T$ \\
\midrule
$1.0\times10^{-5}$ & 3.249 & $2.1\times10^{-4}$ & $1.6\times10^{-3}$ & 0.14 & $4.4\times10^{-1}$ & $4.7\times10^{-1}$ \\
$3.0\times10^{-5}$ & 3.138 & $4.4\times10^{-4}$ & $9.6\times10^{-4}$ & 0.46 & $3.2\times10^{-1}$ & $4.1\times10^{-1}$ \\
$1.0\times10^{-4}$ & 3.019 & $9.6\times10^{-4}$ & $5.6\times10^{-4}$ & 1.71 & $4.3\times10^{-2}$ & $2.4\times10^{-1}$ \\
$2.0\times10^{-4}$ & 2.950 & $1.5\times10^{-3}$ & $4.2\times10^{-4}$ & 3.61 & $1.5\times10^{-4}$ & $8.2\times10^{-2}$ \\
$3.0\times10^{-4}$ & 2.910 & $2.0\times10^{-3}$ & $3.6\times10^{-4}$ & 5.57 & $1.3\times10^{-8}$ & $2.1\times10^{-2}$ \\
$4.7\times10^{-4}$ & 2.865 & $2.7\times10^{-3}$ & $3.0\times10^{-4}$ & 8.97 & $1.5\times10^{-19}$ & $1.1\times10^{-3}$ \\
$1.0\times10^{-3}$ & 2.789 & $4.4\times10^{-3}$ & $2.2\times10^{-4}$ & 19.81 & $<10^{-30}$ & $7.9\times10^{-10}$ \\
\bottomrule
\end{tabular*}
\caption{Linear Langevin predictions along zero-mode trajectories for the reference lattice ($N_{\rm lat}=256$, $L=16000$). For $\gamma-\gc\ge3\times10^{-3}$, $m_H$ exceeds $60$ and both crossing fractions are below $10^{-30}$. The tabulated $m_H$ is the minimum pre-turnaround margin; $\sigma_H$ is evaluated at that minimum.}
\label{tab:lintable}
\end{table*}

\begin{table*}[t]
\centering
\renewcommand{\arraystretch}{1.0}
\setlength{\tabcolsep}{4pt}
\begin{tabular*}{\textwidth}{@{\extracolsep{\fill}}lccccccc@{}}
\toprule
 & & \multicolumn{2}{c}{$\sigma_H^{\max}$} & \multicolumn{2}{c}{$d$} & \multicolumn{2}{c}{$m_H=d/\sigma_H^{\max}$} \\
\cmidrule(lr){3-4}\cmidrule(lr){5-6}\cmidrule(lr){7-8}
Run & $\gamma_{\rm eff}-\gc$ & Theory & Lattice & Theory & Lattice & Theory & Lattice \\
\midrule
fiducial & $9.1\times10^{-5}$ & $5.9\times10^{-4}$ & \ensuremath{6.5\times10^{-4}} & $9.1\times10^{-4}$ & \ensuremath{9.0\times10^{-4}} & 1.55 & \ensuremath{1.38} \\
noise off & $1.3\times10^{-4}$ & $2.3\times10^{-4}$ & \ensuremath{2.5\times10^{-4}} & $1.2\times10^{-3}$ & \ensuremath{1.2\times10^{-3}} & 5.12 & \ensuremath{4.71} \\
vacuum IC & $1.5\times10^{-5}$ & $1.2\times10^{-3}$ & \ensuremath{1.2\times10^{-3}} & $2.9\times10^{-4}$ & \ensuremath{2.8\times10^{-4}} & 0.24 & \ensuremath{0.23} \\
$\Delta t=1$ & $5.0\times10^{-4}$ & $2.9\times10^{-4}$ & \ensuremath{3.1\times10^{-4}} & $2.8\times10^{-3}$ & \ensuremath{2.8\times10^{-3}} & 9.56 & \ensuremath{9.02} \\
$384^3$ & $6.5\times10^{-4}$ & $2.6\times10^{-4}$ & \ensuremath{2.5\times10^{-4}} & $3.3\times10^{-3}$ & \ensuremath{3.3\times10^{-3}} & 12.68 & \ensuremath{13.25} \\
$128^3,\ L=8000$ & $1.1\times10^{-4}$ & $4.3\times10^{-4}$ & \ensuremath{4.5\times10^{-4}} & $1.0\times10^{-3}$ & \ensuremath{1.0\times10^{-3}} & 2.41 & \ensuremath{2.28} \\
\bottomrule
\end{tabular*}
\caption{Linear Langevin predictions and lattice measurements in six near-threshold runs. The effective $\gamma$ is fixed by turnaround; $\sigma_H^{\max}$ is the pre-turnaround maximum and $d$ is evaluated at turnaround.}
\label{tab:fluct}
\end{table*}

\section{Thermodynamic relations and capture criterion}
\label{app:thermo}

\subsection{Thermodynamic relations}

The free-energy density of the bath is $F=-|a_0|(1+rg^2)T^4$, where $T$ is the radiation temperature, $g=g(\Phi)$ is the modulus-dependent coupling, and $r$ sets the strength of its contribution. The entropy density, radiation energy density, and pressure are
\begin{equation}
s=4|a_0|(1+rg^2)T^3,\qquad
\rho_r=3|a_0|(1+rg^2)T^4,\qquad
p_r=\frac{\rho_r}{3}.
\end{equation}
They satisfy $\dd\rho_r=T\,\dd s+(\partial F/\partial\Phi)_T\,\dd\Phi$, so the force on the modulus at fixed entropy is $-(\partial F/\partial\Phi)_T=+\sqrt{2/3}\,a_2g^2T^4$, the thermal force in \eq{eq:eom}. Writing the interaction term as $a_2g^2T^4=-[rg^2/(3(1+rg^2))]\rho_r$ and differentiating at fixed $\rho_r$ instead would give a force larger by the factor $1/(1+rg^2)$, which equals $1.23$ at the initial field value. We therefore use the fixed-entropy force throughout.

The same relations guarantee energy conservation. Without friction and gradients, $\dot\rho_\Phi=-3H\dot\Phi^2-(\partial F/\partial\Phi)_T\dot\Phi$, while conservation of comoving entropy gives $\dot\rho_r=-3H(\rho_r+p_r)+(\partial F/\partial\Phi)_T\dot\Phi$; their sum is the total continuity equation. With friction, the kinetic energy lost at the rate $\eta\dot\Phi^2$ is deposited in the bath as heat, $T\dot s=\eta\dot\Phi^2$, and the noise returns energy to the lattice modes at the rate $\eta T/(a^3\Delta x^3)$ per unit physical volume; these are the two terms of \eq{eq:entropy}. In the lattice implementation, the heat exchanged in each Ornstein--Uhlenbeck update equals the exact change in kinetic energy during that update.

\subsection{Capture criterion for the homogeneous mode}

We now justify the criterion used to classify homogeneous trajectories as captured after turning around on the inner side of the potential barrier. Define the homogeneous field energy as $E=\tfrac12\dot\phi^2+\Veff(\phi,T(t))$ and the instantaneous barrier energy as $V_b(T)=\Veff(\Phibar(T),T)$, where $\Phibar(T)$ is the position of the barrier maximum. Both quantities depend on time because the radiation temperature evolves.

Using the homogeneous form of \eq{eq:eom}, with $\nabla^2\Phi=0$ and $\xi=0$, and the condition $\partial_\Phi\Veff=0$ at the barrier top, we obtain
\begin{equation}
\frac{\dd}{\dd t}\,(E-V_b)=-(3H+\eta)\dot\phi^2+4a_2T^3\dot T\,\big[g^2(\phi)-g^2(\Phibar)\big]\le0.
\end{equation}
The inequality holds while $\dot T\le0$ and $\phi\le\Phibar(T)$. The first term is nonpositive because Hubble friction and dissipation remove kinetic energy. The second term is also nonpositive because $a_2>0$, the temperature is decreasing, and $g^2$ decreases with $\Phi$, so that $g^2(\phi)\ge g^2(\Phibar)$ on the inner side of the barrier.

At a turnaround before the barrier, $\dot\phi=0$ and the field energy lies below the barrier energy, $E<V_b$. Since $E-V_b$ cannot increase while the above conditions hold, the field cannot subsequently reach the barrier top during that interval. This provides the basis for treating a homogeneous trajectory that turns around on the inner slope as captured. The argument assumes $\dot T\le0$, which is satisfied while cosmological expansion dominates the heating from friction, up to and shortly after the turnaround. At later times, oscillations of the captured field can reheat the radiation, so the inequality alone does not establish capture indefinitely.

We therefore continue the zero-mode trajectories for $2\times10^5$ time units after turnaround, about twenty oscillation periods in the well, for $\gamma$ between $\gc+10^{-6}$ and $1$. None returns to the barrier top, and the energy margin $V_b-E$ is smallest at turnaround itself. For individual regions in the lattice simulations, spatial fluctuations and noise invalidate this argument near the separatrix, as the vacuum-initialized realization demonstrates.

\section{Numerical validation}
\label{app:validation}

\begin{figure}[!htbp]
\centering
\includegraphics[width=\linewidth]{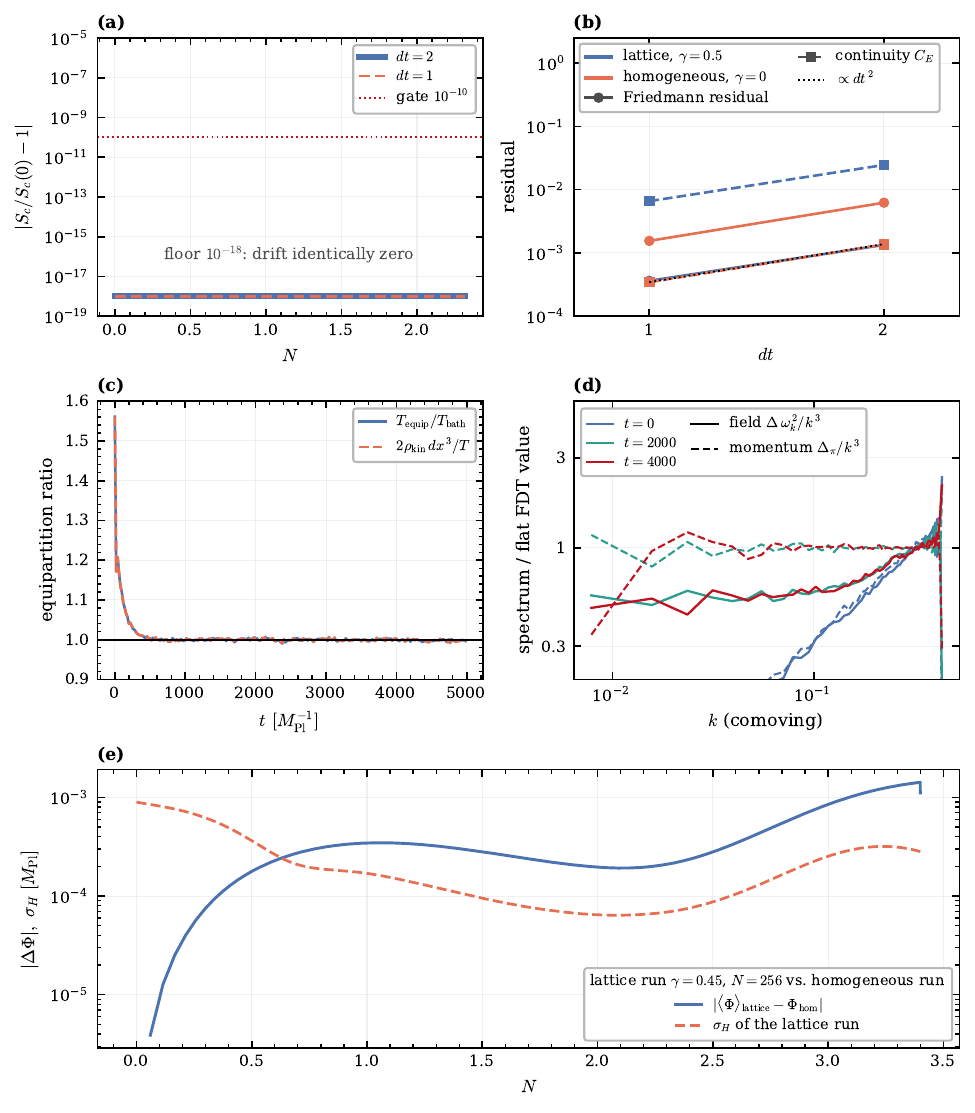}
\caption{Validation tests of the stochastic lattice code. (a) Comoving-entropy conservation. (b) Time-step scaling of the Friedmann and continuity residuals. (c) Kinetic equipartition in Minkowski space. (d) Equilibrium field and momentum spectra. (e) Three-dimensional versus homogeneous evolution and Hubble-smoothed variance.}
\label{fig:validation}
\end{figure}

We validate the lattice implementation through conservation laws, an equilibrium fluctuation--dissipation test, and comparisons with an independent homogeneous solver. The checks are shown in \fref{fig:validation}. When $\eta=0$, the comoving entropy is conserved, giving $Ta/T_i=[(1+rg_i^2)/(1+r\ave{g^2})]^{1/3}$, where $g_i=g(\Phi_i)$ is the initial coupling; the measured drift vanishes identically, against a tolerance of $10^{-10}$, and the closed-system law $T(a)$ at $\gamma=0$ is reproduced to $\ensuremath{9.9\times10^{-13}}$. The residuals of the Friedmann constraint and of the total-energy continuity equation $C_E$ are required to stay below $10^{-2}$ and $5\times10^{-2}$, and both decrease as $\Delta t^2$: halving the time step reduces $C_E$ from $\ensuremath{2.5\times10^{-2}}$ to $\ensuremath{6.6\times10^{-3}}$ in the $\gamma=0.5$ lattice run and from $\ensuremath{1.4\times10^{-3}}$ to $\ensuremath{3.4\times10^{-4}}$ in the $\gamma=0$ homogeneous lattice run. For a free field in Minkowski space with constant $\eta$, the kinetic equipartition temperature $T_{\rm equip}=\mathrm{Var}(\pi)\Delta x^3/a^3$ reaches $\ave{T_{\rm equip}/T}=\ensuremath{0.9992}\pm\ensuremath{2.7\times10^{-3}}$, while the total energy of field and bath drifts by $\ensuremath{-1.5\times10^{-5}}$.

Of the 41\ runs in the original campaign, two exceed the specified tolerances: the $T^3$-friction run has a Friedmann residual of $\ensuremath{1.2\times10^{-2}}$, and a $\gamma=0.52$ repeat with coarser output has a continuity residual $C_E=\ensuremath{8.5\times10^{-2}}$, where $C_E$ denotes the residual of the total-energy continuity equation. Neither run enters the near-threshold analysis. All other runs, including the extended and convergence runs, satisfy the tolerances; for the latter runs the Friedmann residual is at most $4.5\times10^{-3}$ and $C_E\le2.5\times10^{-2}$. The comoving horizon also remains within half the periodic box size, reaching $\ensuremath{6.3\times10^{-1}}$ of $L/2$ in the original campaign and at most $0.77$ when the $L=8000$ convergence runs are included. This horizon condition does not apply to the Minkowski test, which has $H=0$ and is evolved for several light-crossing times.

\begin{figure}[!htbp]
\centering
\includegraphics[width=\linewidth]{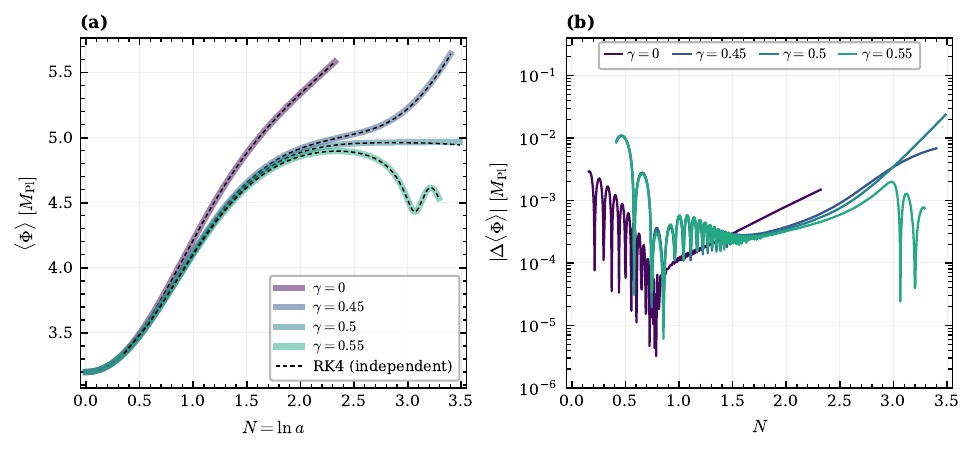}
\caption{Independent homogeneous cross-check. (a) Field evolution from the noise-free lattice code (solid) and the separate RK4 zero-mode solver (dashed). (b) Absolute difference between the two, including the enhanced near-threshold deviation at $\gamma=0.50$.}
\label{fig:crosscheck}
\end{figure}

As an independent cross-check, we run the lattice code without stochastic noise or initial fluctuations and compare it with an RK4 integration of the homogeneous equations using a separate code, which shares only the model definition and uses a time step of $0.5$. The two calculations agree over their common integration interval away from the capture threshold (\fref{fig:crosscheck}). The near-critical case $\gamma=0.50$ is more sensitive: the independent zero mode turns around at $\Nturn=2.865$, whereas the lattice trajectory with $\Delta t=2$ remains close to the barrier and has passed it by $2.6\times10^{-3}$ at $t=3\times10^4$. This difference results from a time-step shift of $-5.3\times10^{-4}$ in the effective dissipation parameter (\aref{app:offsets}), amplified by the logarithmic sensitivity of $\Nturn$ near the separatrix.

\begin{figure}[!htbp]
\centering
\includegraphics[width=\linewidth]{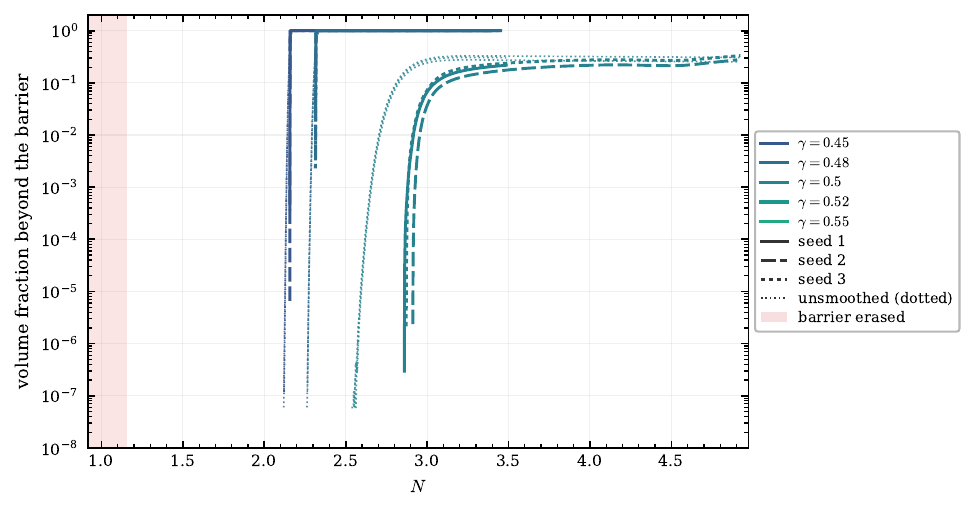}
\caption{Barrier-crossing fractions in the reference scan at nominal $\gamma=0.45$, $0.48$, $0.50$, $0.52$, and $0.55$: Hubble-smoothed field (solid) and unsmoothed field (dotted). The shaded region marks the interval without a thermal barrier. Line styles distinguish noise realizations; fractions below one lattice site are omitted.}
\label{fig:fcross}
\end{figure}

\Fref{fig:fcross} shows the barrier-crossing fractions in the reference scan for the Hubble-smoothed and the unsmoothed field. In the overshot runs at nominal $\gamma\le0.48$, the whole volume passes the barrier together with the average field, and both fractions reach unity. In the runs at nominal $\gamma=0.52$ and $0.55$, no lattice site crosses the barrier. Only the run at nominal $\gamma=0.50$, which lies within the stochastic layer, shows a partial crossing, with maximum fractions of $0.22$ for the Hubble-smoothed field and $0.31$ for the unsmoothed field. The two observables therefore agree on which runs cross the barrier locally, while the value of the fraction depends on the filter.

To classify the evolution of the spatially averaged field, we record its first turnaround, where $\ave{\dot\Phi}$ changes sign from positive to negative while $\ave\Phi<\Phibar(T)$. A run is classified as overshot if, at any output after the barrier has reappeared, $\ave\Phi$ exceeds $\Phibar(T)+0.05\,[\Phibar(T)-\Phimin(T)]$. For a run that has turned around, we estimate at the last output the remaining coasting distance $\Delta\Phi_{\rm rem}=2\ave{\dot\Phi}/[3(1-w_{\rm tot})H]$, where $w_{\rm tot}$ is the total equation-of-state parameter, and classify the run as captured if $\Delta\Phi_{\rm rem}<\Phibar(T)-\ave\Phi$ and as undecided otherwise. This expression neglects direct dissipation ($\eta=0$); for $\eta=\gamma H$, its denominator becomes $3(1-w_{\rm tot})+2\gamma$, reducing the remaining coasting distance and making the stated capture test conservative. No separate settling-time or energy-based condition is used.

The test is informative only when the average field moves outward at the last output. If it moves inward, $\Delta\Phi_{\rm rem}<0$ and the run is counted as captured, which then means only that the average field has turned around below the barrier and has not passed it by the end of the run. Because the estimate also neglects the inward potential force between $\Phimin$ and $\Phibar$, it overestimates the outward excursion: a run that passes the test while moving outward is expected to remain inside the barrier during that excursion, whereas a failed test need not indicate overshoot. In the reference scan, only the $\gamma=1$ run ends moving outward and passes the test, while the runs at $\gamma=0.50$, $0.55$, $0.60$, and $0.75$ end moving inward. The $\gamma=0.52$ run ends on its second outward swing ($N=3.38$) and fails the test, with $\Delta\Phi_{\rm rem}=1.26$ against a remaining distance of $0.38$, although it lies $0.02$ above the zero-mode threshold and no lattice site crosses the barrier. The classification of the average field is therefore a finite-time statement, as the vacuum-initialized realization illustrates: it is classified as captured at $t=3\times10^4$ and crosses the barrier when continued (\sref{sec:fate}). For the zero mode we use the energy argument of \aref{app:thermo} instead.

The late-time evolution of the barrier-crossed regions in \sref{sec:fate} is more sensitive to resolution. At the reference spacing $\Delta x=62.5$, the characteristic interface width $|\Veff''|^{-1/2}$ corresponds to $1.4$ and $1.1$ lattice cells at $N=3.9$ and $4.1$, when the crossing fraction approaches its plateau, and decreases to $0.55$ cells by $N=4.8$. The unstable band and the Hubble-smoothed field remain resolved up to approximately $N=4.4$, while the later interface structure becomes increasingly sensitive to the lattice resolution. We therefore extend a run at $\Delta x=31.25$ ($N_{\rm lat}=256$, $L=8000$), where the interface spans $2.7$ and $2.2$ cells at $N=3.9$ and $4.1$. At $\gamma_{\rm eff}-\gc=7.9\times10^{-5}$, its plateau crossing fraction is $0.28$. The fraction beyond the outer inflection point increases from $0.10$ at $N=4.5$ to $0.18$ at $N=4.7$ and $0.23$ at $N=4.8$, and the maximum Hubble-smoothed field reaches $\Phi_H=5.92$. As in $s_2$, the average field begins to increase again near the end of the run.

The Ornstein--Uhlenbeck map of \eq{eq:OU} is exact for fixed $a$, $T$, and $\eta$, but the full split evolution includes an evolving bath, nonlinear forces, and energy exchange. Its second-order time-step bias is measured by the runs at $\Delta t=2,1,0.5$ and separated from the near-threshold fluctuation effects. These tests establish the numerical accuracy claimed within the discretized classical model.

\section{Near-threshold numerical offsets}
\label{app:offsets}

\begin{figure}[!htbp]
\centering
\includegraphics[width=\linewidth]{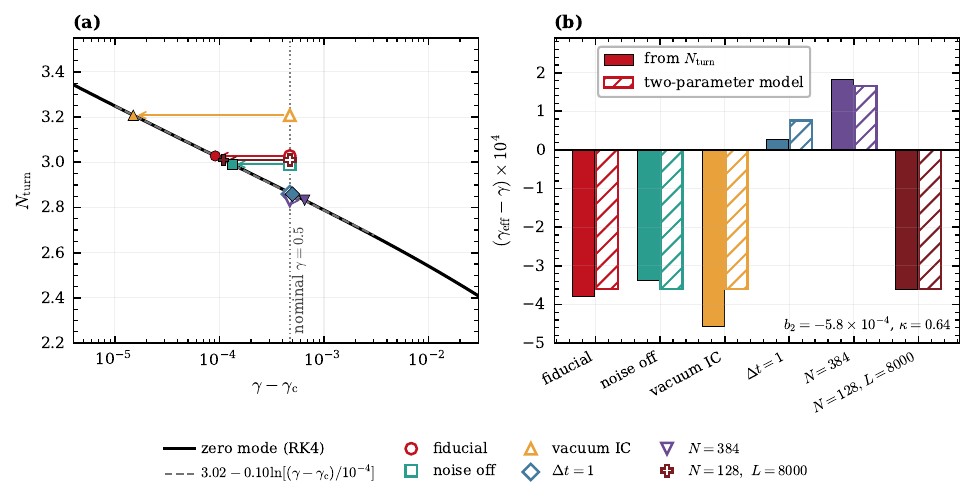}

\caption{Effective dissipation parameter near the capture threshold. (a) Zero-mode turnaround curve (solid), logarithmic fit (dashed), and lattice results at nominal (open) and matched effective (filled) $\gamma$; arrows indicate the matching procedure. (b) Measured offsets (solid bars) and a two-parameter fit (hatched) based on \eq{eq:offmodel}, with $X=-\kappa\,\Delta\gc(\Delta x)$, where $\Delta\gc$ is the threshold shift estimated by including $\rho_{\rm lat}$ in the zero-mode Friedmann equation. Fitting the five thermal-spectrum checks gives $b_2=-5.8\times10^{-4}$ and $\kappa=0.64$. The independently determined time-step coefficient is $b_2=-5.31\times10^{-4}$, obtained from homogeneous lattice convergence tests. The separate calibration used for the convergence ensemble is described in the text.}
\label{fig:effgamma}
\end{figure}

Near the capture threshold, the turnaround time is highly sensitive to small changes in the dynamics because of the logarithmic relation in \eq{eq:Nturnlog}. We therefore define an effective dissipation parameter $\gamma_{\rm eff}$ for each lattice realization: it is the value of $\gamma$ for which the independent zero-mode integration reproduces the lattice turnaround time $\Nturn$. This matching uses only the trajectory of the spatially averaged field, not any fluctuation observable. The difference $\gamma_{\rm eff}-\gamma$ measures the effective shift relative to the input (nominal) dissipation parameter. The mapping from lattice turnaround times to $\gamma_{\rm eff}$ and the resulting offsets are illustrated in \fref{fig:effgamma}.

Two numerical effects dominate this shift. First, the second-order time integration introduces a contribution proportional to $\Delta t^2$. Second, the thermalized lattice modes carry an additional energy density $\rho_{\rm lat}\simeq T/(a\Delta x)^3$, which changes the expansion rate and hence the capture threshold. At the initial time, $\rho_{\rm lat}/\rho_r$ is $2.25\times10^{-4}$, $1.80\times10^{-3}$, and $6.07\times10^{-3}$ for $N_{\rm lat}=128$, $256$, and $384$, respectively, at fixed $L=16000$. We describe the measured offset by
\begin{equation}
\gamma_{\rm eff}-\gamma
=b_2\Big(\frac{\Delta t}{2}\Big)^2+X(\Delta x),
\label{eq:offmodel}
\end{equation}
where $b_2$ is the time-step offset at $\Delta t=2$ and $X(\Delta x)$ is the contribution associated with the lattice modes at spacing $\Delta x$. This decomposition provides a practical calibration of the discretized trajectories.

We isolate the time-step term using a homogeneous lattice with $N_{\rm lat}=16$, no noise, and no initial fluctuations, so that no thermalized fluctuation modes contribute. Bisecting its capture threshold to $10^{-6}$ gives values above $\gc$ by $5.31\times10^{-4}$, $1.33\times10^{-4}$, $3.33\times10^{-5}$, and $7.9\times10^{-6}$ for $\Delta t=2$, $1$, $0.5$, and $0.25$, respectively. Their successive ratios are $3.99$, $3.99$, and $4.2$, consistent with second-order convergence. Hence $b_2=-5.31\times10^{-4}$, and the homogeneous discretized threshold approaches $\gc$ as $\Delta t\to0$ (\fref{fig:convergence}a).

Subtracting the time-step contribution from the convergence runs gives $X(62.5)=(1.65\pm0.37)\times10^{-4}$ from 17 runs at $\Delta t=2$, $(1.56\pm0.19)\times10^{-4}$ from four runs at $\Delta t=1$, and $(1.1\pm0.4)\times10^{-4}$ from three runs at $\Delta t=0.5$, where the uncertainties are standard deviations across runs. These measurements support the additive form of \eq{eq:offmodel} to within approximately $5\times10^{-5}$, comparable to the scatter between realizations. At smaller lattice spacings we find $X(41.7)=(7.2\pm0.2)\times10^{-4}$ and $X(31.25)=(1.89\pm0.01)\times10^{-3}$. Over this range, $X$ grows approximately as $\Delta x^{-3.6}$, somewhat faster than $\rho_{\rm lat}\propto\Delta x^{-3}$. Including $\rho_{\rm lat}$ in the zero-mode Friedmann equation reproduces the sign and order of magnitude of the shift, but not its full value, because it neglects the pressure and energy exchange of the lattice modes.

For the larger convergence ensemble, the offsets were instead calibrated separately for each pair $(\Delta t,\Delta x)$ before predicting the new runs (\tref{tab:convergence}). Across the 23 runs that turned around, the predicted $\gamma_{\rm eff}$ agrees with the measured value to an r.m.s.\ error of $3.4\times10^{-5}$, with a maximum deviation of $1.2\times10^{-4}$. The first group at $\Delta t=0.5$, whose offset was not yet calibrated, remained near the barrier top without turning around; it was repeated at $\gamma=0.49953$ with the offset inferred from that group, giving three turnarounds within $7\times10^{-5}$ of the prediction. This repeated group is not counted among the 23 predicted runs.

\begin{table*}[!t]
\centering
\begin{tabular}{@{}cccccccc@{}}
\toprule
$N$ & $\Delta t$ & $\Delta x$ & $\gamma$ & \multicolumn{2}{c}{$(\gamma_{\rm eff}-\gc)\times10^5$} & \multicolumn{2}{c}{plateau of $\fcross^{H}$}\\
\cmidrule(lr){5-6}\cmidrule(lr){7-8}
 & & & & predicted & measured & lattice & theory\\
\midrule
128 & 2 & 62.5 & 0.49979 & $-10.1$ & overshot & 0.66--0.76 & --\\
128 & 2 & 62.5 & 0.49994 & $+4.9$ & 3.5, 1.4, 3.9 & 0.40, 0.46, 0.38 & 0.40, 0.46, 0.38\\
128 & 2 & 62.5 & 0.49998 & $+8.9$ & 20.5, 10.1, 10.1 & 0.06, 0.21, 0.23 & 0.07, 0.23, 0.23\\
128 & 2 & 62.5 & 0.50004 & $+14.9$ & 9.1, 11.5, 11.4 & 0.26, 0.19, 0.12 & 0.25, 0.21, 0.21\\
128 & 2 & 62.5 & 0.50019 & $+29.9$ & 28.9, 30.7, 31.2 & 0.0005, 0.003, 0.004 & 0.02, 0.01, 0.01\\
128 & 2 & 62.5 & 0.50049 & $+59.9$ & 62.0, 57.1, 57.9 & $<10^{-4}$ & $<10^{-4}$\\
128 & 1 & 62.5 & 0.49959 & $+8.9$ & 8.3, 10.9, 5.5 & 0.25, 0.19, 0.35 & 0.27, 0.22, 0.34\\
128 & 0.5 & 62.5 & 0.49943 & $+8.6$ & no turnaround & 0.50--0.57 & --\\
128 & 0.5 & 62.5 & 0.49953 & $+10.0$ & 10.0, 3.3, 11.3 & 0.24, 0.38, 0.21 & 0.24, 0.40, 0.21\\
192 & 2 & 41.7 & 0.49944 & $+9.3$ & 11.5, 6.7, 11.9 & 0.25, 0.27, 0.24 & 0.21, 0.31, 0.20\\
256 & 2 & 31.25 & 0.49824 & $+8.5$ & 7.9, 5.9 & 0.28, 0.31 & 0.28, 0.33\\
\bottomrule
\end{tabular}
\caption{Near-threshold convergence runs with $L=8000$, thermal initial fluctuations, and noise. Predicted and measured threshold distances are compared with the lattice and linear-theory crossing plateaus (median over $3.9<N<4.4$). ``Overshot'' and ``no turnaround'' indicate the outcomes of the spatially averaged field.}
\label{tab:convergence}
\end{table*}

The physical conclusions do not depend on separating the offset into these two contributions. The calibration explains why identical nominal values of $\gamma$ can correspond to different distances from the capture threshold and allows the lattice fluctuation results to be compared with linear Langevin theory on a homogeneous background matched to the spatially averaged lattice field.

\FloatBarrier

\bibliographystyle{apsrev4-2}
\bibliography{refs}

\end{document}